\documentclass[aps]{revtex4-2}

\usepackage{bbold}
\usepackage[colorlinks=true, allcolors=blue]{hyperref}
\usepackage{graphicx}
\usepackage{braket}
\usepackage{enumerate}
\usepackage{amsmath,amsfonts,amssymb,amsthm}
\usepackage{nicefrac}
\usepackage{bm}
\usepackage{mathrsfs}
\usepackage{comment}
\usepackage{color}
\usepackage{dsfont}
\usepackage[capitalise]{cleveref}
\usepackage{algorithm, algpseudocode}

\DeclareSymbolFont{largesymbolsA}{U}{txexa}{m}{n}
\DeclareMathSymbol{\varprodb}{\mathop}{largesymbolsA}{16}
\newcommand{\varprod}{\mathrel{\scalebox{1.2}{$\varprodb$}}}

\algblockdefx[qIf]{qIf}{qEndIf}
[2][]{\textbf{\underline{if}} #2 \textbf{\underline{then}}}
{\textbf{\underline{end} \underline{if}}}

\algcblockdefx[qIfElse]{qIf}{qElse}{qEndIf}
{\textbf{\underline{else}}}
{\textbf{\underline{end} \underline{if}}}

\algnewcommand{\IfThen}[2]{\State \algorithmicif\ #1\ \algorithmicthen\ #2}
\algnewcommand{\qIfThen}[2]{\State \textbf{\underline{if}}\ #1\ \textbf{\underline{then}}\ #2}
\algnewcommand{\cqIfThen}[2]{\State \textbf{if}\ #1\ \textbf{\underline{then}}\ #2}

\crefformat{section}{\S#2#1#3}

\theoremstyle{definition}
\newtheorem{remark}{Remark}

\newtheorem{definition}{Definition}
\newtheorem{quo}{Quotation}
\newtheorem{example}{Example}
\newtheorem{task}{Task}

\begin{document}

\title{Framework for Learning and Control in the Classical and Quantum Domains}

\author{Seyed Shakib Vedaie}
\altaffiliation{Corresponding author.}
\altaffiliation{These authors contributed equally.}
\email{seyedshakib.vedaie@ucalgary.ca}
\affiliation{Institute for Quantum Science and Technology, University of Calgary, 2500 University Drive NW, Calgary, Alberta, T2N 1N4, Canada}

\author{Archismita Dalal}
\altaffiliation{These authors contributed equally.}
\email{archismita.dalal1@ucalgary.ca}
\affiliation{Institute for Quantum Science and Technology, University of Calgary, 2500 University Drive NW, Calgary, Alberta, T2N 1N4, Canada}

\author{Eduardo J.~P\'aez}
\email{eduardo.paez1@ucalgary.ca}
\affiliation{Institute for Quantum Science and Technology, University of Calgary, 2500 University Drive NW, Calgary, Alberta, T2N 1N4, Canada}

\author{Barry C.~Sanders}
\email{sandersb@ucalgary.ca}
\affiliation{Institute for Quantum Science and Technology, University of Calgary, 2500 University Drive NW, Calgary, Alberta, T2N 1N4, Canada}

\date{\today}
\begin{abstract}
Control and learning are key to technological advancement, both in the classical and quantum domains,
yet their interrelationship is insufficiently clear in the literature, especially between classical and quantum definitions of control and learning.
We construct a framework that formally relates learning and control,
both classical and quantum,
to each other,
with this formalism showing how learning can aid control.
Furthermore, our framework helps to identify interesting unsolved problems in the nexus of classical and quantum control and learning
and helps in choosing tools to solve problems.
As a use case,
we cast the well-studied problem of adaptive quantum-enhanced interferometric-phase estimation as a supervised learning problem
for devising feasible control policies.
Our unification of these fields relies on diagrammatically representing the state of knowledge,
which elegantly summarizes existing knowledge and exposes knowledge gaps.

\end{abstract}

\maketitle

\section{Introduction}
\label{introduction}

Closed-loop control aims to regulate a system of devices to produce a desired output autonomously~\cite{DB08,RZB18},
and machine learning (ML) is about improving a procedure autonomously through experience~\cite{Mit97}.
ML can be subdivided into reinforcement learning (RL) and semi-supervised learning, with supervised learning (SL) and unsupervised learning (UL) the extreme cases~\cite{RN20}.
Both control and learning are vital to advancing technology and are interrelated:
learning can assist with enhancing control
(such as the field of `machine-learning control')~\cite{DBN17},
and control tools could be used for improving learning~\cite{EHL19}.
Adding to the richness of this field,
both control and learning can be cast in a classical
(i.e., non-quantum) context~\cite{DB08,MBW+19}
and in a quantum context~\cite{WM09,Jac14,ZLW+17,BWP+17}.
These fields are rapidly developing with concomitant problems of inconsistent terminology, disparities between classical and quantum versions,
and currently less-than-clear relations between these four topics:
classical and quantum control and learning,
i.e., four areas of classical control ($\mathscr{C}$C),
quantum control ($\mathscr{Q}$C),
classical learning ($\mathscr{C}$L) and quantum learning ($\mathscr{Q}$L).
To be succinct, we refer to these topics as `learning for control', and our aim is to construct a unified framework connecting these topics, which we call our learning-for-control framework (LfC).
To this end, we summarize state-of-the-art in connecting these topics together and show the value of our LfC framework by 
how it reveals knowledge gaps and by showing how applying SL
to quantum-enhanced metrology~\cite{HS11,LCPS13}
reveals a procedure for performing this task autonomously.

We begin by summarizing state of the art,
and we represent this state of the art by constructing a knowledge graph~\cite{HBC+21}
whose vertices are current topic areas,
and both directed and undirected edges show connections between topic areas.
Furthermore, knowledge graph edges are labelled by the actual references.
This knowledge graph is particularly useful to convey not just state of the art but also to convey knowledge gaps.

Next, we unify $\mathscr{C}$C and $\mathscr{Q}$C.
Although both of these topics are focused on control, specifically closed-loop control,
the literature for $\mathscr{Q}$C differs, even from basic construction,
from the literature for $\mathscr{C}$C.
$\mathscr{C}$C is founded on concepts such as a controller,
the controller's policy for controlling a plant,
a reference used to achieve a particular output state, and feedback~\cite{DB08}.
$\mathscr{Q}$C typically focuses on techniques to optimize quantum systems by controlling coefficients in a Hamiltonian or open-system evolution~\cite{WM09,Jac14}.
We formulate a new version of $\mathscr{Q}$C that essentially quantizes $\mathscr{C}$C,
which is distinct from our earlier approach of defining $\mathscr{Q}$C independent of $\mathscr{C}$C~\cite{VPS18}.

Building on this $\mathscr{C}$C-$\mathscr{Q}$C unification,
we explore unifying $\mathscr{C}$L and $\mathscr{Q}$L in the same way.
Our work sets the stage for this unification by revisiting foundational aspects of $\mathscr{C}$L but considers whether the basic objects therein can be treated mathematically.
Specifically,
we articulate desiderata for $\mathscr{Q}$L based on extending $\mathscr{C}$L concepts and based on unification commensurate with our $\mathscr{C}$C-$\mathscr{Q}$C analysis.

Our next step is to extend the idea of learning for control~\cite{Fu70} to the quantum domain.
This extension involves allowing for quantum channels connecting the controller, plant and learner.
Furthermore, the no-cloning principle of quantum mechanics~\cite{Wil17} forbids the same feedback to be sent, accurately and deterministically, to both learner and controller,
which we account for in our framework.

We base our approach on Fu's framework for learning control systems~\cite{Fu70}.
We explicitly separate teacher, user, learner and controller in our framework, as distinct from Fu's approach, which equates controller and learner and does not include `user' in the picture (hence, not complete).
In our LfC framework, the teacher implements the process of learning for control for RL and semi-supervised learning, which differs from Fu's approach, where the teacher is only present for the SL case.

We demonstrate the utility of our LfC framework in casting adaptive quantum-enhanced interferometric-phase estimation  (A$\mathscr{Q}$P)~\cite{HS11,LCPS13}
as a SL problem.
The benefit of introducing SL into this framework is that new circumstances -- such as modified working conditions -- can be accommodated by using the SL model to predict new possibilities rather than having to optimize for each case,
which is intrinsically harder.
A$\mathscr{Q}$P is a form of quantum-enhanced metrology,
useful for enhancing quantum clocks~\cite{BS13}
and interferometric position-shift measurements~\cite{Holl79,CTD+80,Cav81} inter alia,
but with feedback incorporated,
yielding the advantage that only single-particle measurements are required rather than joint measurements if adaptive feedback methods are not used.

Beyond the practical application of using our framework for methodically applying learning to control,
our work is interesting on a philosophical level.
Does control theory make sense for a quantum controller,
for example, a controller with a quantum computer and with quantum information coming to or leaving the controller?
Could the controller prepare the plant in a superposition of states or be entangled with the plant?
Could the controller have a superposition of policies?
We do not have the answers to such questions,
but one value of our work is that such questions arise very clearly from our framework.

Our paper continues in~\S\ref{sec:background} with a review of the key concepts in ML, control theory, unification of classical and quantum mechanics,
and learning for A$\mathscr{Q}$P.
We then elaborate on how we construct our LfC framework in~\S\ref{sec:framework}.
Based on our literature survey and the key concepts established in~\S\ref{sec:framework}, we present our knowledge graph in \S\ref{sec:graphrep}.
Next, in~\S\ref{sec:casting}, we describe the application of our framework to A$\mathscr{Q}$P. 
Finally, we discuss our results in~\S\ref{sec:discussion}
and conclude in~\S\ref{sec:conclusion} with a summary of our work and an outlook.
As we employ many abbreviations, we summarize these abbreviations and their full expressions in Table~\ref{tab:glossary} for the convenience of the reader.
\begin{table}[]
\centering
\begin{tabular}{c|l}
\textbf{Abbreviation} & \textbf{Description} \\
\hline
ML & Machine learning \\
$\mathscr{C}$C & Classical control \\
$\mathscr{Q}$C & Quantum control \\
$\mathscr{C}$L & Classical learning \\
$\mathscr{Q}$L & Quantum learning \\
SL & Supervised learning \\
UL & Unsupervised learning \\
RL & Reinforcement learning \\
$\mathscr{Q}$ML & Quantum machine learning \\
POVM & Positive operator-valued measure \\
LfC & Learning-for-control \\
A$\mathscr{Q}$P & Adaptive quantum-enhanced\\ & interferometric-phase estimation \\
 
\end{tabular}
\caption{Glossary}
\label{tab:glossary}
\end{table}

\section{Background}
\label{sec:background}

In this section, we give a review of the concepts of $\mathscr{C}$C and ML, along with their quantum counterparts.
We further provide a summary of the concepts of learning for control, both in  
classical and quantum domains. 
The essential background on the $\mathscr{Q}$C technique of A$\mathscr{Q}$P is also provided.

\subsection{Machine learning}
\label{subsec:ML}

ML is proving to be highly valuable due to its capacity for making predictions based on experience.
We begin by discussing the framework for ML.
Then we briefly explain classical ML,
which includes the widely used modes of SL and UL
as well as RL.
Finally, we discuss relevant definitions and notions of quantum ML ($\mathscr{Q}$ML).

\subsubsection{Framework for machine learning} 
\label{subsubsec:frameworklearning}

Now we discuss the essentials of ML.
First, we define ML and then differentiate learning from optimization. Finally, we explain the classifications of an ML task.

We begin by establishing the concept of ML and discussing its extension to the quantum case.
We adopt and formalize Mitchell's definition of ML.
\begin{quo}[Mitchell~\cite{Mit97}]
\label{quote:clearning1}
A computer program is said to learn from experience~$\mathscr{E}$ with respect to some class of tasks~$\mathscr{T}$ and performance measure~$\mathscr{P}$, if its performance at tasks in~$\mathscr{T}$, as measured by~$\mathscr{P}$, improves with experience~$\mathscr{E}$.
\end{quo}
\noindent
Russell and Norvig formalize the concept of a computer program in the context of ML by introducing the notion of an agent.
\begin{quo}[Russell \& Norvig~\cite{RN20}]
\label{quote:rn_agent}
An agent is just something that acts \dots. Of course, all computer programs do something, but computer agents are expected to do more: operate autonomously, perceive their environment, persist over a prolonged time period, adapt to change, and create and pursue goals.
\end{quo}
\noindent
Quotations~\ref{quote:clearning1} and~\ref{quote:rn_agent} clearly highlight the essential components of an ML implementation, namely learning agent~$\mathscr{A}$, $\mathscr{E}$, $\mathscr{T}$ and~$\mathscr{P}$,
which we regard as essential constructs for $\mathscr{Q}$ML as well.

A model is a mathematical representation of the data,
specifically,
a formula or algorithm that labels the data in a probably approximately correct way.
The goal of ML is to search for a model that optimizes the performance of a learning agent~\cite{Mit97,MRT18}.
Various notions of representation exist, so here we give our definition, pertinent throughout this paper.
\begin{definition}
\label{def:representation}
A representation of a function~$f(x)$
is~$\tilde{f}_d(x)$,
with~$d$ specifying the size (e.g., number of bits) and $\|f-\tilde{f}_d\|$
the distance between the function and its representation
with respect to the norm on function space.
\end{definition}
\begin{example}
\label{example:representations}
A truncated Taylor expansion~\cite{Rud76} is one example of a representation.
\end{example}
\begin{example}
\label{example:cumulant}
Another example of a representation of a function is a cumulant expansion~\cite{van2007stochastic}
with the sequence of cumulants denoted
\begin{equation}
\label{eq:cumulants}
    \bm\varpi := \left(\kappa_1, \kappa_2, \kappa_3, \ldots\right),
\end{equation}
for~$\kappa_1$ the mean, $\kappa_2$ the variance and $\kappa_3$ the skewness.
\end{example}

ML is classified based on the nature of~$\mathscr{E}$.
For example, we can consider what we call the structure of~$\mathscr{E}$ -- labelled or unlabelled features as one pair of cases
or state-and-reward as another case -- 
as a way of classifying;
alternatively,
we can consider all~$\{\mathscr{E}\}$
being provided up-front or is simultaneous with the ML process.
ML can be categorized into three major 
paradigms based on the structure of~$\mathscr{E}$,
namely SL
for~$\mathscr{E}$ being labelled features,
UL for unlabelled features
and 
RL for~$\mathscr{E}$ being state-and-reward~\cite{SB18}.
If all~$\{\mathscr{E}\}$ is provided up-front,
ML is offline;
ML is online if~$\mathscr{E}$
involves input data from the system in 
real time while learning.
\begin{figure}[t!]
\centering   
\includegraphics[width=0.55\columnwidth]{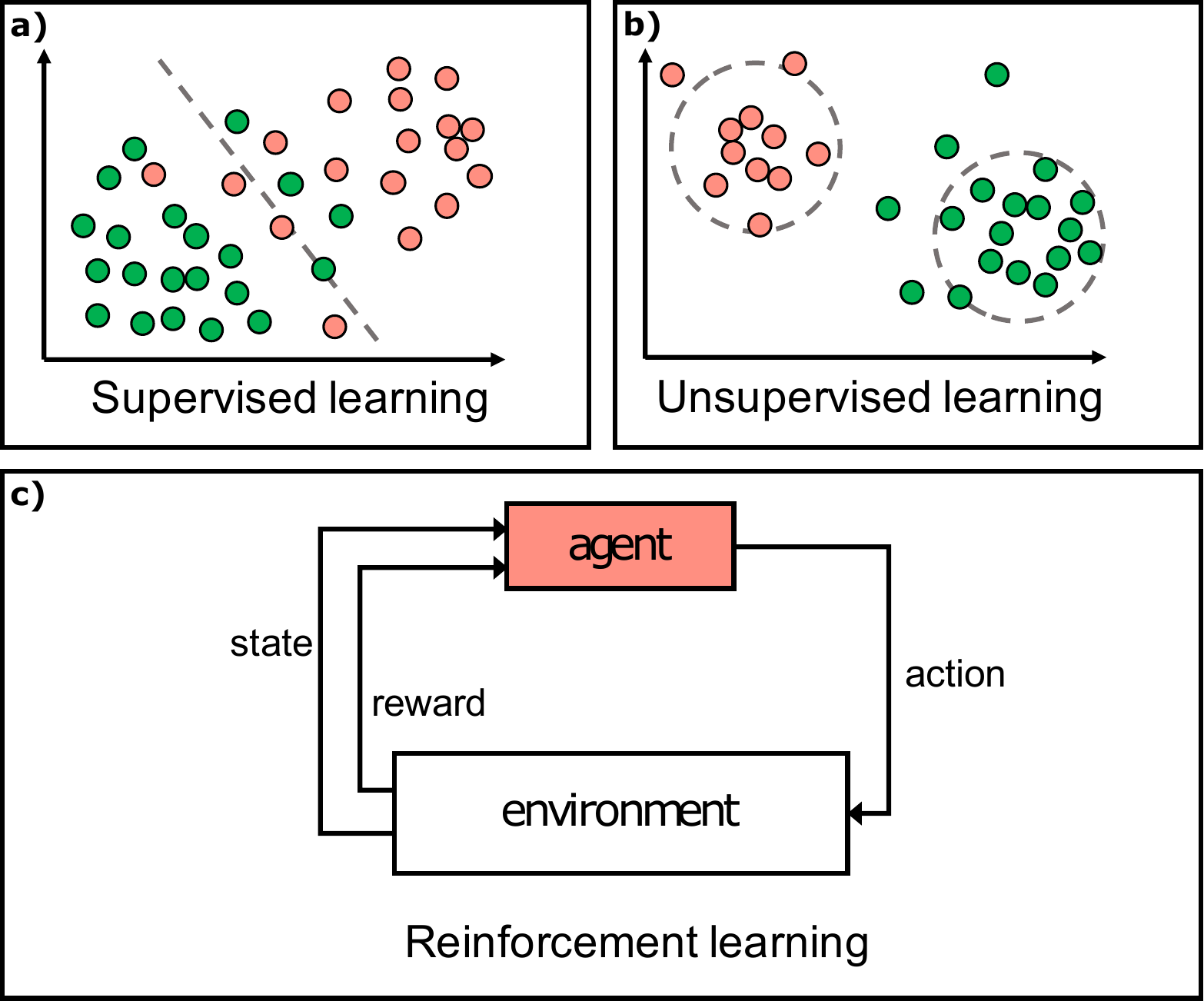}
\caption{Machine learning classification. (a) An example of the supervised learning paradigm showing a decision boundary that classifies input data. (b) An example of the unsupervised learning paradigm that represents the clustering of input data. (c) A typical workflow in the reinforcement learning paradigm that demonstrates how learning proceeds via interactions between an agent and its environment.}
\label{fig:ml_paradigms}
\end{figure}

Every ML algorithm comprises three components---representation, evaluation and optimization~\cite{Dom12}. The representation component concerns the hypothesis space of the learner~\cite{Mit97,MRT18}.
The evaluation component concerns the objective function for learning and the learning performance. Finally, the optimization component concerns the methods to search for a model in the hypothesis space that maximizes the learning performance. Thus, optimization adds value to all ML problems by enhancing ML, and optimization is not learning in itself. 

\subsubsection{Machine learning paradigms}
\label{subsubsec:classicalmachinelearning}

Now, we elaborate on three major paradigms of ML and discuss a typical ML workflow.
We begin by summarizing the ML pipeline, which entails four steps.
These four steps are preprocessing raw data into a training data set, followed by the second step of training, followed by the validation step, and finally, the testing step.
We then discuss SL and UL briefly yet sufficiently to establish their mutual differences and the comparative difference with RL, which we elaborate on at the end.

We describe an ML pipeline as an iterative workflow with four steps typically, namely 
\begin{equation}
\label{eq:ml_pipeline}
\text{pre-processing} \rightarrow \text{calibrating} \rightarrow \text{training} \rightarrow  \text{testing}.
\end{equation}
The workflow begins with the data pre-processing step, where raw data is manipulated to construct a data set~$\mathscr{D}$ that is suitable for ML. The pre-processing step involves several sub-tasks such as data cleansing, feature extraction and feature selection~\cite{AOOD20}.
The pre-processed data set is then divided into the disjoint union of a calibrating and training set~$\mathscr{D}_\text{model}$ and a testing set~$\mathscr{D}_\text{test}$ as
\begin{equation}
\label{eq:Ddisjointunion}
    \mathscr{D}=\mathscr{D}_\text{model}\sqcup \mathscr{D}_\text{test}.
\end{equation}
In the calibration step, the hyperparameters of the model are tuned in an iterative
way~\cite{Bur89}. 
For each tuple of hyperparameters, a model
is trained on a randomly-sampled subset~$\mathscr{D}_\text{train}$ of~$\mathscr{D}_\text{model}$, and the model's performance is
evaluated on the remaining data set. A mean performance, corresponding to each tuple of hyperparameters,
is then calculated by repeating these two sub-steps for different~$\mathscr{D}_\text{train}$. After
repeating this process of calculating mean performance for all possible tuples of hyperparameters,
the calibration step returns the tuple corresponding to the best model performance.
In the training step, the model data set~$\mathscr{D}_\text{model}$, along with the hyperparameters returned
from the calibration step, are used to construct an ML model.
Finally, the model,
with selected parameters and hyperparameters, is assessed on~$\mathscr{D}_\text{test}$.
The model passes or fails at the test step;
if the model passes, this model is then used on real data.

In the SL paradigm~\cite{RN20},
using Mitchell's terminology~\cite{Mit97},~$\mathscr{E}$ is provided to~$\mathscr{A}$ as a labelled data set~$\mathscr{D}$ of size~$\mathscr{N}$.
For each labelled data being a tuple of an $F$-dimensional feature vector~$\bm{x}_i$ and its corresponding label~$y_i$, we can formally state the data set, according to a convenient formalism~\cite{DBS21}, as
\begin{equation}
\label{eq:sl_dataset}
    \mathscr{D}=\{(\bm{x}_i,y_i)
    \mid i\in [\mathscr{N}]:=\{0,\ldots, \mathscr{N}-1\}\} 
    \subset \mathbb{R}^F \times \mathbb{R}.
\end{equation}
Theoretically,
we allow real-valued entries,
but computationally,
we approximate by floating-point numbers up to machine precision.
~$\mathscr{A}$ then devises a labelling map
\begin{equation}
\label{eq:sl_map}
f:\mathbb{R}^F\to\mathbb{R}:
    \bm{x}\mapsto \tilde{y},
\end{equation}
where~$\bm{x}$ is an unseen feature vector and~$\tilde{y}$ is its corresponding predicted label. 
Labelling is not guaranteed to be correct every time but it rather is probabilistically approximately correct~\cite{KV94,Val84}. Denoting the set of all labels as~$Y$, the fitness of~$f$ is quantified by a `loss function'
\begin{equation}
    L:Y\times Y \mapsto \mathbb{R}^+,
\end{equation}
that measures the difference between~$y$ and~$\tilde{y}$. Common examples of loss functions include the absolute and the squared losses~\cite{KRF09}.
SL tasks are further sub-classified into two types: classification and regression~\cite{MRT18}.
A classification task is defined for discrete labels, whereas regression is pertinent for the case of continuous labels.

In the case of UL, the agent has access to unlabelled input data~$\{\bm{x}_i\}$
but not to labels~$\{y_i\}$.
The task of a UL agent is to recognize hidden patterns in the data set and to use these patterns to cluster these data together into subsets~\cite{MRT18}.
In practical applications,
SL and UL can be combined to form semi-supervised learning~\cite{MRT18}, which is especially useful for tasks involving training data with few input-label pairs and thus a large number of input-only data.

RL is quite different from SL and UL:
instead of having labelled or unlabelled input data,
respectively,
RL involves an agent, states of the environment, actions and rewards.
These notions are subtle and inconsistently defined
so we continue our practice of relying on verbatim quotes from respected sources to define these entities.
The environment plays a key role for RL,
with the environment described in the following quote.
\begin{quo}[Russell \& Norvig~\cite{RN20}]
\label{quote:rl_env}
The environment could be everything -- the entire
universe! In practice it is just that part of the 
universe whose state we care about when
designing this agent -- the part that affects what the 
agent perceives and that is affected by
the agent's actions.
\end{quo}
\noindent
In RL, the agent performs actions on the environment that are intended to yield the best outcome,
and the agent is rewarded accordingly for good actions.

The agent acquires information from the environment and acts on this information to change the environment.
The nature of the agent is explained in the following quotation.
\begin{quo}[Russell \& Norvig~\cite{RN20}]
\label{quote:agent}
An agent is anything that can be viewed as perceiving its environment through sensors and
acting upon that environment through actuators.
\end{quo}
\noindent
To be clear on how actuators and sensors are defined,
we provide the following quotations.
\begin{quo}[Dorf \& Bishop~\cite{DB08}]
An actuator is a device employed by the control system to alter or adjust the environment.
\end{quo}
\begin{quo}[Dorf \& Bishop~\cite{DB08}]
A sensor is a device that provides a measurement of a desired external signal.
\end{quo}\noindent
The agent is not just reacting to the environment to modify its state but rather learns how to improve the environment as Sutton and Barto explain.
\begin{quo}[Sutton \& Barto~\cite{SB18}]
\label{quote:reward}
Reinforcement learning is learning what to do -- how to map situations to actions -- so as to maximize a numerical reward signal.
\end{quo}
\noindent
Russell and Norvig elaborate on the agent's nature.
\begin{quo}[Russell \& Norvig~\cite{RN20}]
\label{quote:best}
[A]gents are expected to \dots\ operate
autonomously, perceive their environment, persist over a prolonged time period, adapt to
change, and create and pursue goals.
A rational agent is one that acts so as to achieve the
best \dots expected outcome.
\end{quo}
\noindent
Sutton and Barto articulate the ``prolonged time period'' in Quotation~\ref{quote:best}
as the agent
``maximizing not the immediate reward,
but cumulative reward in the long run''~\cite{SB18}.
The ``reward signal'' in Quotation~\ref{quote:reward}
and the ``best outcome'' in Quotation~\ref{quote:best}
rely on an agent whom we regard as hidden.

This reward, or ``reward signal''~\cite{SB18},
which formalizes ``the purpose or goal of the agent''~\cite{SB18} and is passed ``from the environment to the agent''~\cite{SB18},
has an origin that transcends the agent and the environment as two entities.
The reward
is vitally important to ``formaliz[ing] the idea of a goal''~\cite{SB18}
and seems to demand that a third party come into play.
Sutton and Barto say:
\begin{quo}[Sutton \& Barto~\cite{SB18}]
The reward signal is your way of communicating to the agent what you want achieved.
\end{quo}
\noindent
What does ``your way of communicating to the agent'' mean?
This statement implies subjectivity by another agent, namely you.
The reward thus represents the ``creator'' of the problem who 
imbues subjective values on how actions should be rewarded.
\noindent
\begin{quo}[Sutton \& Barto~\cite{SB18}]
\label{quote:rl_and_evolution}
Our focus is on reinforcement learning methods that learn while interacting with the environment, which evolutionary methods do not do. \dots Evolutionary methods ignore much of the useful structure of the reinforcement learning problem: they do not use the fact that the policy they are searching for is a function from states to actions; they do not notice which states an individual passes through during its lifetime, or which actions it selects. \dots Although evolution and learning share many features and naturally collaborate, we do not consider evolutionary methods by themselves to be especially well suited to reinforcement learning problems \dots
\end{quo}

\subsubsection{Online versus offline learning}

ML paradigms can be employed in both online and offline settings. First, we define the concepts of online and offline learning. Then, we treat the concept of RL in both online and offline settings separately as RL inherently assumes online interaction with an environment.

We begin by explaining offline ML,
which is applied for the case that all~$\mathscr{E}$ is available before ML starts.
Offline ML then searches for a feasible model by learning the parameters of the model by using the entire training data set for~$\mathscr E$,
which is fully available at the outset.

In contrast to offline ML,
online ML is employed while the training set of~$\mathscr{E}$ becomes available;
ideally, online ML acts immediately as each element of~$\mathscr E$
is received rather than storing elements~\cite{BKY97}.
For example, in online SL,
~$\mathscr{A}$ has to predict the label of the next instance based on the labels of the previous instances that~$\mathscr{A}$ has already seen. Online learning is helpful if training over the entire data set is computationally infeasible or  the algorithm must adapt to new patterns in the data dynamically.
The following quote conveys the distinction between offline and online learning.
\begin{quo}[Ben-David, Kushilevitz \& Mansour~\cite{BKY97}]
The difference between the models is that, while in the on-line model only the set of possible elements is known, in the off-line model the sequence of elements (i.e., the identity of the elements as well as the order in which they are to be presented) is known to the learner in advance. 
\end{quo}

Now we consider online vs.~offline RL.
Per the construction of RL,
which involves~$\mathscr A$ interacting with the environment discussed in~\ref{quote:rl_env},
RL is implicitly online.
However, real-time collection of experience via interaction with an environment could be expensive or dangerous~\cite{LKTF20,PMC23}.
To avoid such problems,
offline RL is an important topic.
Offline RL accommodates agents who utilize data from previously collected experience,
which could have been obtained from simulation or for a safe setting.
In contrast to offline SL and UL,
which are well posed and widely used,
reconciling RL,
which is implicitly online,
with offline methods is a work in progress, as emphasized in the following quote.
\begin{quo}[Levine, Kumar, Tucker \& Fu~\cite{LKTF20}]
[O]ffline RL is, at its core, a counter-factual
inference problem: given data that resulted from a given set of decisions, infer the consequence of a \textit{different} set of decisions. Such problems are known to be exceptionally challenging in machine learning, because they require us to step outside of the commonly used i.i.d.\ framework, which assumes that test-time queries involve the same distribution as the one that produced the training data.
\end{quo}

\subsubsection{Quantum machine learning}
\label{subsubsec:qenhml}

We review definitions of $\mathscr{Q}$ML as stated explicitly in the literature.
Then, we explain the concept of  $\mathscr{Q}$ML as gleaned from these somewhat disparate definitions.
Finally, we provide some examples of  $\mathscr{Q}$ML.

A few comprehensive review articles~\cite{BWP+17, DB18,CHI+18} and textbooks~\cite{SP21,Wit14} provide various subjective definitions of  $\mathscr{Q}$ML.
Unlike classical ML,
a compact and unanimous definition for  $\mathscr{Q}$ML in terms of the learning components~$\mathscr{A}$, $\mathscr{E}$, $\mathscr{P}$ and $\mathscr{T}$
(in Quotation~\ref{quote:clearning1})
is lacking.
As essential background,
we provide key quotations from some of the most notable literature in this field.
In an early review paper,
a functional definition of  $\mathscr{Q}$ML, without hinting at the prospect of ``quantum advantage''~\cite{Pre12,Pre18,HM17,CHI+18},
follows~\cite{SSP15}.
\begin{quo}[Schuld, Sinayskiy \& Petruccione~\cite{SSP15}]
\label{quote:qlearning1}
    In quantum machine learning, quantum algorithms are developed to solve typical problems of machine learning using the efficiency of quantum computing. This is usually done by adapting classical algorithms or their expensive subroutines to run on a potential quantum computer.
\end{quo}
\noindent
More recently,  $\mathscr{Q}$ML definitions embed the concept of quantum advantage.
 $\mathscr{Q}$ML is defined in a well-cited survey paper as follows.
\begin{quo}[Biamonte et al.~\cite{BWP+17}]
\label{quote:qlearning2}
    The field of quantum machine learning explores how to devise and implement quantum software that could enable machine learning that is faster than that of classical computers.
\end{quo}
\noindent
In a recent perspective article on this field,
Wiebe defines  $\mathscr{Q}$ML from a data-science and computer-science perspective as follows.
\begin{quo}[Wiebe~\cite{Wie20}]
\label{quote:qlearning3}
Quantum ML involves using a quantum device to solve a machine learning task with greater speed or accuracy than its classical analogue would allow.
\end{quo}
\noindent
Thus, we see that  $\mathscr{Q}$ML is defined in three ways:
by its utility as in Quotation~\ref{quote:clearning1},
by the aspiration of  $\mathscr{Q}$ML as in Quotation~\ref{quote:qlearning2}
and by the  $\mathscr{Q}$ML device as in Quotation~\ref{quote:qlearning3}.

Notable examples of  $\mathscr{Q}$ML are varied.
One example is a quantum support-vector machine, where a quantum matrix-inversion algorithm solves the system of equations 
that appear in a least-squares formulation of the support-vector machine~\cite{RML14}.
Another example of $\mathscr{Q}$ML is the analogue-quantum kitchen sink,
which quantizes the classical random kitchen sinks algorithm.
The classical random kitchen sink algorithm estimates a kernel by using randomized features
and can be enhanced by employing adiabatic quantum evolution to yield requisite randomness~\cite{NVS+20}.
A third example is quantum-enhanced RL, for which quantum speedups in the agent's decision-making process and learning times have been claimed~\cite{PDM+14,SAH+21}.

\subsection{Control System}
\label{subsec:control}

In this subsection, we summarize studies of control.
We begin with a summary of $\mathscr{C}$C.
Then we discuss the topic of $\mathscr{Q}$C.
Finally, we summarize the status of applying ML to control.

\subsubsection{Classical control system}
\label{subsubsec:CC}

Now, we present the essentials of $\mathscr{C}$C.
First, we discuss the task of a control system.
Then, we present a schematic for $\mathscr{C}$C that represents the essential elements and their connections.
Finally, we discuss the nature of a control policy and how this policy is devised.

A control task involves steering specific controllable degrees of freedom of a physical system such that its dynamics yields the desired observations within a required tolerance. 
This task is achieved by a control system, which can be defined in one of the following two ways, as examples.
\begin{quo}[Dorf \& Bishop~\cite{DB08}]
\label{quote:dorf_ccontrol}
    A control system is an interconnection of components forming a system configuration that will provide a desired system response.
\end{quo}
\noindent
An alternative definition is the following.
\begin{quo}[Rosolia et al.~\cite{RZB18}]
\label{quote:rosolia_ccontrol}
    A control system is a device in which sensed quantities are used to generate an autonomous behaviour through computation and actuation.
\end{quo}
\noindent
Control systems are classified into 
closed-loop and open-loop control according to whether control depends on feedback or not,
respectively.
\begin{quo}[Dorf \& Bishop~\cite{DB08}]
A closed-loop system uses a measurement of the output signal and a
comparison with the desired output to generate an error signal that is used
by the controller to adjust the actuator.
\end{quo}

\begin{figure}
\centering
\includegraphics[width=0.47\columnwidth]{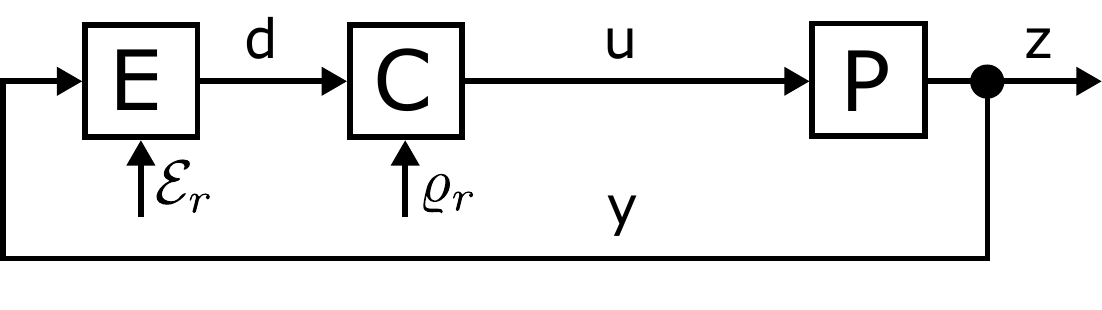}
\hspace{5mm}
\includegraphics[width=0.33\columnwidth]{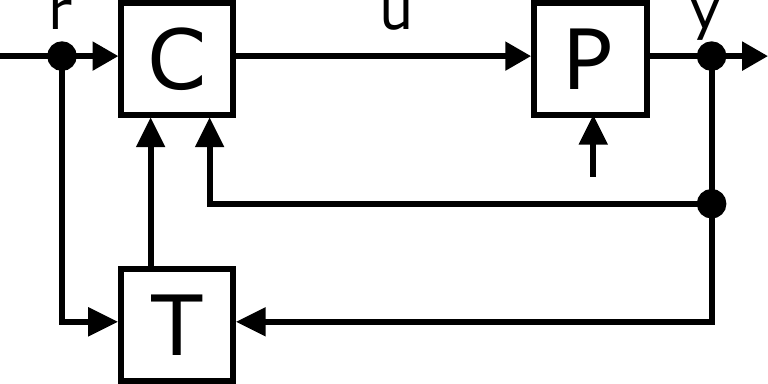}\\
\hspace{10mm} (a) \hspace{80mm} (b)
\caption{(a) A block diagram of a closed-loop $\mathscr{C}$C system, where the arrows between two blocks represent the flow of classical information.
A plant~P, comprising the physical system, actuators and sensors, is controlled by a controller~C who sends a control signal~u to P.
The filled circle beside P represents a ``switching" operation that either fed back P's output into the control system as a signal~y or passes it on as the ultimate output signal~z, which typically contains the termination signal and observable representing the final state of~P. The switching operation in $\mathscr{C}$C is implemented as a fan-out.
An evaluator~E estimates an error signal~d between y and the reference signal~r using an evaluation function~$\varepsilon_\text r$.
C computes~u based on a control policy~$\varrho_\text r$, which is designed based on~r, and~d.
In a setting where y=$\emptyset$, one obtains an open-loop control system.
(b) A block diagram representing the existing learning-for-control framework~\cite{Fu70}.
Here C is a learning controller that devises~$\varrho$ using the information~y from P such that the control goal~r is satisfied. 
The teacher~T evaluates the performance of~C and directs the learning performed by~C such the system's overall performance is gradually improved. 
~T may or may not be involved in the learning loop depending upon whether the learning is done following a SL or UL scenario~\cite{Fu70}.}
\label{fig:control_back}
\end{figure}

We represent a generic closed-loop control system as a block diagram
in Fig.~\ref{fig:control_back}(a). Control systems consist of two main parts: the controlled object and the controller. Our block diagram explains the essential components of closed-loop control.

\begin{remark}
\label{remark:card}
The sets of all actuator data, all possible plant output and all error signals are
\begin{equation}
\label{eq:UY}
U:=\{u\},\,
Y:=\{y\},\,
D:=\{d\},
\end{equation}
with each element~$u$ being distinct from all other elements in~$U$ and, similarly,
each $y$ being distinct from all other elements in~$Y$.
Moreover, the sets~$U$, $Y$ and $D$ have constant cardinalities~$|U|$, $|Y|$ and~$|D|$, respectively, over successive feedback loops.

\end{remark}

\begin{remark}
In the classical case,
each channel is labelled by a letter such as~u (for actuator data),
written in upright (e.g., Roman) font,
with its value in some instance given by (slanted, or italic, font)
$u \in U$,
with (capital letter)~$U$ being the set of all possible (control) values.
The probability vector for the state of~u is
\begin{equation}
\label{eq:pvector}
\bm{p}^{\text{u}}:=\left(p^{\text{u}}_u\right)
\end{equation}
with the right-hand side being the sequence of probability-vector components,
which adds to~$1$.
\end{remark}

\begin{remark}
The quantum state of a channel,
such as~u,
is given by a density operator~\cite{Wil17}, which is a positive semi-definite trace-class operator on Hilbert space,
and a classical probability distribution is recovered by diagonalizing the given operator~\cite{Zur81}.
\end{remark}

A policy is a set of instructions that determine the control parameters, and hence the effectiveness of the control scheme.
In closed-loop feedback,
the controller executes a policy
\begin{equation}
\label{eq:policy}
\varrho:\mathbb{Z}\times D\to U:
    (r,d)\mapsto u.
\end{equation}
with~$r$ representing the value of the reference channel~r. Contrariwise, in an open-loop control scheme, where~y is either not available or is unnecessary, C steers~P to obtain~z without ever receiving any feedback from~P.

Now we explain two standard methods to construct~$\varrho$, namely conventional model-based control (MBC) methods and alternative data-driven control (DDC) methods.
\begin{quo}[Hou \& Wang~\cite{HW13}]
Data-driven control includes all control theories and methods in which the controller is designed by directly using on-line or off-line I/O data of the controlled system or knowledge from the data processing but not any explicit information from mathematical model of the controlled process, and whose stability, convergence, and robustness can be guaranteed by rigorous mathematical analysis under certain reasonable assumptions. 
\end{quo}
\noindent
MBC proceeds by modelling the plant either using first principles or identification from data \cite{HW13} and assumes a trusted mathematical model governing plant dynamics with bounded modelling uncertainty.
The plant's model is represented by the map
\begin{equation}
\label{eq:plantdynamics}
\varsigma:U\to Y:u\mapsto y
\end{equation}
with resources
(e.g., fresh water, battery energy, oxygen) implicitly consumed through this mapping
as elements of~$R$
are not typically stated in standard references.
Then~$\varrho$ is devised based on~$\varsigma$ with the belief that the model is a trustworthy approximation of the true system.
If~$\varsigma$~(\ref{eq:plantdynamics})
is either unknown or not readily solved analytically, then DDC methods can help.
For DDC, $\varrho$ is exclusively devised based on input-output data taken from~P and thus lacks systematic design-and-analysis tools.

\subsubsection{Quantum control system}
\label{subsubsec:quantumcontrol}

Now we discuss $\mathscr{Q}$C.
First, we review disparate definitions of $\mathscr{Q}$C as stated explicitly in the literature. Then, we explain the concept of $\mathscr{Q}$C extracted from these somewhat disparate definitions.

In contrast to $\mathscr{C}$C, a concise definition of $\mathscr{Q}$C is lacking. Therefore, we provide two quotations from notable literature on 
$\mathscr{Q}$C.
\begin{quo}[Walmsley \& Rabitz~\cite{WR03}]
\label{quote:qcontrol1}
    Quantum control refers to active intervention in a system’s dynamics to maximize the probability, based on a given metric, that the system evolves toward a desired target state.
\end{quo}
\begin{quo}[Lloyd~\cite{Llo00}]
\label{quote:qcontrol2}
In the conventional picture of quantum feedback control, sensors perform measurements on the system, a classical controller processes the results of the measurements, and actuators supply semiclassical potentials to alter the behaviour of the quantum system.
\end{quo}
\noindent
Although the two quotations appear to be different, these quotations are compatible in that the probability in Quotation~\ref{quote:qcontrol1} is inferred from sampling by sensors in Quotation~\ref{quote:qcontrol2} and the active intervention in Quotation~\ref{quote:qcontrol1} is elaborated as classical control and altering semi-classical potentials in Quotation~\ref{quote:qcontrol2}.
Neither of these quotations is sufficiently clear on what components of a $\mathscr{Q}$C system are classical or quantum.

Now we explain the techniques for constructing~$\varrho$ in $\mathscr{Q}$C, which have traditionally taken an MBC approach. The standard practice in $\mathscr{Q}$C is to construct a first-principle model~$\varsigma$ that describes the evolution of~P over time, with~$u$ manipulating one or more coefficients of the Hamiltonian describing~P~\cite{BCS17,BCR10,ZLW+17}. Then~$\varrho$ is devised based on~$\varsigma$ via gradient-based greedy algorithms. 
A popular greedy optimization algorithm is gradient-ascent pulse engineering, which was first proposed to obtain~$\varrho$ for control tasks in nuclear magnetic resonance spectroscopy~\cite{KRC+05}. 
Although the fitness landscape for~$\varrho$ in $\mathscr{Q}$C is often compatible with greedy algorithms, sometimes greedy algorithms yield poor results, especially for constrained large-dimensional quantum systems~\cite{ZSS14}.
Therefore, MBC methods employing global-optimization algorithms, such as evolutionary algorithms~\cite{ES15}, have been developed to obtain~$\varrho$ for $\mathscr{Q}$C problems such as quantum-enhanced adaptive phase estimation~\cite{HS11,LCPS13} and quantum gate design~\cite{ZGS15,ZGS16,ZSS14}.
For the cases where~$\varsigma$ is either unknown or not readily solved analytically~\cite{BCR10}, ideas from DDC $\mathscr{C}$C have been used in $\mathscr{Q}$C with success~\cite{TGB15,BPB16,AN17,WEH+16,MGCC15,GK10}. However, the lack of a formalized structure prevents the rapid progress of DDC $\mathscr{Q}$C techniques.

\subsubsection{Learning for control}
\label{subsubsec:learn}

We now proceed to discuss literature applying ML techniques for both types of control, i.e., classical and quantum.
We begin by summarizing the concept of machine-learning control.
Then we summarize the state of the art for classical ML applied to both~$\mathscr{C}$C and~$\mathscr{Q}$C.
Finally, we summarize the literature on $\mathscr{Q}$ML for~$\mathscr{C}$C.

Machine-learning control is the concept of using ML algorithms to learn an effective~$\varrho$ for a control system. This concept is motivated by control problems where identifying and modelling a plant is challenging or infeasible due to unobservability and highly non-linear effects. Machine-learning control typically involves a ``learning controller'' who employs learning techniques to execute a control task~\cite{Fu70, DBN17}. Learning enables a controller, who is neither omniscient nor possesses a feasible alternative, to execute the task successfully by assisting in designing feasible control policies. 
ML methods are highly flexible and adaptable but lack systematic design-and-analysis tools for studying their stability and robustness in the context of control~\cite{HW13}.
\begin{remark}
To cast a control problem into a learning problem, we identify the components of a control task as the components of learning, namely, $\mathscr{A}$, $\mathscr{T}$, $\mathscr{P}$ and~$\mathscr{E}$.
~$\mathscr{A}$ devises~$\varrho$ for~C such that the control task is achieved. In this scenario~$\mathscr{E}$ comprises plant feedback and control signal. The user chooses~$\mathscr{P}$, which is calculated based on~r.
\end{remark}

Classical ML has been widely applied for controlling classical operations.
We highlight three approaches to incorporating ML into $\mathscr{C}$C.
(i)~Using ML to construct~$\varsigma^{-1}$ without prior knowledge of~$\varsigma$~(\ref{eq:plantdynamics})~\cite{DBN17,HW13}.
(ii)~Using ML for system identification, i.e., learning~$\varsigma$ based on training data. The obtained~$\varsigma$ is then used to design~$\varrho$~(\ref{eq:policy}) following MBC methods~\cite{MRH18,BLN+22}. 
(iii)~Using ML for safe control, where a learning agent uses data to safely improve performance by learning the uncertain dynamics~\cite{BGH+22}.
Fu's seminal work~\cite{Fu70} introduces a framework for utilizing $\mathscr{C}$L for $\mathscr{C}$C, where an agent called teacher~(T) assess and directs learning of a classical~C towards better performance
as shown in Fig.~\ref{fig:control_back}(b).

Classical ML has been applied to $\mathscr{Q}$C as well.
Classical ML and heuristic optimization techniques are actively used for designing high-performance quantum gates for fault-tolerant quantum computing~\cite{AZ19}.
Additionally, neural networks are used in some other $\mathscr{Q}$C tasks including adaptive quantum tomography and dynamic decoupling~\cite{AN17}. 
Meta-heuristic algorithms are also shown to be advantageous in designing $\mathscr{Q}$C policies that are robust to random noise and decoherence~\cite{PS19}.

We now consider the application of $\mathscr{Q}$ML for control. 
 $\mathscr{Q}$ML for $\mathscr{C}$C has been studied in the context of quantum RL~\cite{MUP+22}.
RL is a valuable tool for solving direct adaptive optimal control problems~\cite{SBW92}. Therefore, developing quantum variations of RL algorithms is an important topic. Recent works on the concept of quantum RL, however, are limited to solving simple $\mathscr{C}$C problems such as the problem of maze traversal being solved on a quantum annealer~\cite{JGM+21,SJD22,CLG+18,LCG+17,CYQ+20,LS20,LS21,SWIK22}.

An intriguing new intersection between $\mathscr{Q}$C and  $\mathscr{Q}$ML is Variational Quantum Algorithms (VQAs)~\cite{CAB+21,BCK+22,TCC+21}. The relationship between VQAs and $\mathscr{Q}$C is grounded in that they can both be seen as quantum-classical optimization tasks, which facilitates using a shared language between both settings~\cite{MAG+21,GWR22,YRS+17}. In VQAs, a parameterized quantum circuit, often termed an ansatz, is adjusted iteratively using a classical optimization routine and feedback loop to find optimal parameters for a specific task. $\mathscr{Q}$C, on the other hand, aims at steering quantum dynamics for realizing desired states or unitary transformations, often necessitating intricate control sequences. VQAs can be employed in $\mathscr{Q}$C to devise feasible control sequences that achieve a desired control task~\cite{DBC22,SCXC22} or to develop $\mathscr{Q}$ML methods for solving $\mathscr{Q}$C problems~\cite{WJW+23,SSB23}. Recent developments in VQAs have also led to new design-and-analysis tools for control~\cite{HK21,BC21,LPQ+22,LJG+23}.
$\mathscr{Q}$C tools have, in turn, improved the design and analysis of VQAs~\cite{MGB+21,dKTK23,LWC+22,AACA22,IMR+22,MRBA16,YRS+17,GWR22,LCS+22,MAG+21,CDB+21}. For example, $\mathscr{Q}$C techniques have led to better ansatz for VQAs~\cite{CDB+21} and helped diagnose barren plateaus~\cite{LCS+22}.

\subsection{Unifying classical and quantum mechanics}
\label{subsec:classicalquantum}

In this subsection, we explain our approach for combining classical and quantum descriptions of any physical system,
and this framework is important to our overall aim of unifying classical and quantum control and learning.
We begin with a brief review of C$^*$-algebra, which provides a consistent framework for observables independent of whether the system is classical or quantum.
Then, we discuss the correspondence principle and, finally, describe the representation of system states for classical and quantum mechanics.

We adopt an operational viewpoint,
which is about states being descriptions of how a system is prepared, and measurements being about describing detection ~\cite{Bri27,BGL97}.
Formally,
this description can be made rigorous by treating states as linear functionals on the appropriate vector space for the given physics
(classical or quantum with particular symmetries)
and measurement as a positive operator-valued measure (POVM)~\cite{HZ11}.
A duality exists between states and measurements,
and the basic mathematical description is achieved by employing observables within a C$^*$ algebra~\cite{Str08}.
Each observable in this algebra is self-adjoint and is a vector-space homomorphism
(``linear operator'').
The vector space for classical physics is probability space and, for quantum mechanics,
is Hilbert space
or its extension to generalized functions for infinite-dimensional space~\cite{Bal14,Bow20};
here we maintain the simpler language of referring to Hilbert space whether or not the space is finite- or infinite-dimensional,
with the technicalities being covered by literature on the Gel'fand triple~\cite{Mad05,BB15,GG07}.
Evolution is described by mappings known as channels.

We now explain classical and quantum channels based on Wilde's two descriptions~\cite{Wil17},
which we present here as definitions. The following definition rephrases slightly Wilde's description of a classical channel.
\begin{definition}[Wilde~\cite{Wil17}]
\label{def:c_channel}
A classical channel is a conditional probability distribution from input random variable~$X$ to output random variable~$Y$.
\end{definition}
\noindent
The next definition is verbatim from Wilde's book~\cite{Wil17}.
\begin{definition}[Wilde~\cite{Wil17}]
\label{def:q_channel}
A quantum channel is a linear, completely positive, trace-preserving map.
\end{definition}
\noindent
As our objective is to describe learning and control agnostically,
i.e., without regard to whether the object is quantum or classical, we need to combine Defs.~\ref{def:c_channel}
and~\ref{def:q_channel} into a single agnostic definition of a channel:
\begin{definition}
\label{def:channel}
A channel is a linear, completely positive, trace-preserving map acting on an inner-product space.
\end{definition}

Classical-quantum correspondence is a method either for quantizing a classical system or dequantise a quantum system and can be achieved in various ways~\cite{AE05,Lan17}.
One way to achieve classical-quantum correspondence is via geometric correspondence.
Geometry in the classical case arises from the Poisson bracket
with states being represented on phase space,
which is a Poisson manifold~\cite{Arn13}.
Quantum states,
on the other hand,
can be represented on a K\"{a}hler manifold with commutators imbuing a geometry that is analogous to the classical case~\cite{AS95,Ash99,BH01}.
This geometric correspondence can be challenging to establish in a mathematically rigorous way, but we do not require such sophisticated methods here~\cite{Lan17}.
In practice,
as observed by Dirac~\cite{Dir81},
Poisson brackets of observables are simply replaced by commutators with $\text{i}:=\sqrt{-1}$ introduced as a scalar coefficient. With these notions in mind, we can understand that combining classical and quantum definitions of channel into~\ref{def:channel} is consistent with Wilde's claim below.
\begin{quo}[Wilde~\cite{Wil17} \S4.6.4]
\label{quote:classical=quantum_channel}
[C]lassical channels are special cases of quantum channels.
\end{quo}
\noindent
Thus,
if this inner-product space is Hilbert space, then the channel is quantum in concordance with Definition~\ref{def:q_channel}.
If, on the other hand,
the inner-product space is restricted to being probability space
(as a subspace of Hilbert space~\cite{FG13} (ch 20)), then the channel is classical,
matching Definition~\ref{def:c_channel}.

Subsequent to the channel the system is subject to measurement, which we now explain mathematically. Rigorously, measurement corresponds to a POVM which we now describe classically and quantumly.
\begin{definition}[van Fraassen~\cite{vFr08}]
[M]easurement is an operation that locates an item (already classified as in the domain of a given theory) in a logical space (provided by the theory to represent a range of possible states or characteristics of such items).
\end{definition}
\noindent
In the quantum case, every ideal measurement corresponds to an observable, which is a self-adjoint operator in the C$^*$-algebra and, for finite-dimensional Hilbert space, can be represented by a Hermitian matrix. More generally, for blurry measurement, the POVM assigns some probability to each observable. Classically, the observables act on a probability space, and one way to obtain the classical POVM is to diagonalize the matrices representing observables~\cite{Zur81}.

We represent states of classical systems as elements of~$L^1(\mathbb{R})$,
which are square-integrable
real-valued functions over real numbers and their limits, such as Dirac~$\delta$.
Uncertainties are associated with the spread of these distributions,
such as standard deviation~\cite{Moo50},
entropic uncertainty~\cite{CBTW17}
or Cram\'{e}r-Rao lower bound~\cite{Kay93}.
In quantum mechanics, 
uncertainty arises from Born's rule,
for which the distribution of measurement outcomes arises by associating this distribution with the squared modulus of the wavefunction,
itself being a representation of the state in $L^2(\mathbb{R})$
 (complex-valued normalizable functions over real numbers)~\cite{Lan09}.

\subsection{Learning for adaptive quantum-enhanced interferometric-phase estimation}
\label{sec:AQEM}

In this subsection, we summarize work to date on ML methods applied to A$\mathscr{Q}$P.
First, we explain the topic of A$\mathscr{Q}$P.
Then we discuss A$\mathscr{Q}$P in the context of $\mathscr{Q}$C with particular attention to policies,
i.e., control procedures,
and, finally, we elaborate on learning methods for such policies.

\subsubsection{Adaptive quantum-enhanced interferometric-phase estimation}
\label{subsubsec:adaptiveqem}

We now introduce the concept of A$\mathscr{Q}$P and its mathematical description.
First, we focus on the interferometric transformation and the nature of the input state.
We then use this mathematical description to quantify the imprecision in terms of the Holevo variance.  
\begin{figure}[t!]
\centering
\includegraphics[width=0.55\columnwidth]{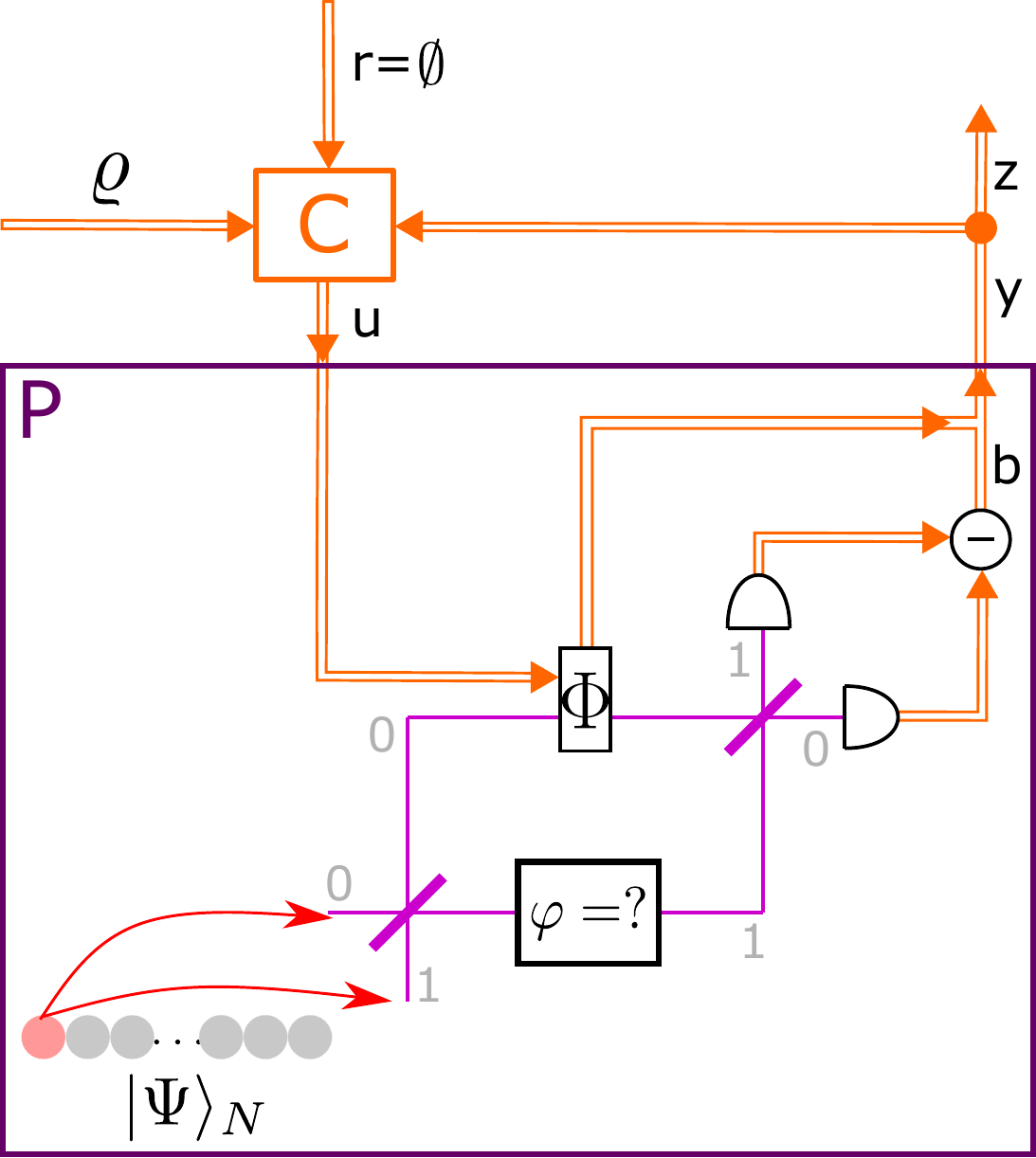}
\caption{A block diagram representing the A$\mathscr{Q}$P scheme as a closed-loop control~[\cref{fig:control_back}(a)]. The scheme utilizes a passive linear lossless two-channel interferometer, such Mach--Zehnder, which has a pair of input and output ports, one element of each pair labelled~0 and the other~1. The interferometer arms are also labelled~0 and~1. Within the interferometer, arm~0 undergoes a known controllable phase shift of~$\Phi$, and arm~1 undergoes an unknown phase shift of~$\varphi$. 
The quantum~P incorporates the interferometer and its accompanying sensors and actuators and is represented by a purple block. The classical~C is represented by an orange block, and the double-lined orange arrows represent the flow of classical information.~C is in charge of sending control signal~$u$ to update~$\Phi$. The update in~$\Phi$ is performed based on a pre-established~$\varrho$ and~P's feedback~$y$, which consists of the photon measurement information bit~$b$ and the current value of~$\Phi$.~y goes to~z as a fan-out, and  
after all~$N$ photons have been injected into the interferometer and measurements are completed, the final bit string for all measurements maps to~$\tilde\varphi$.}
\label{fig:aqp}
\end{figure}

The aim of A$\mathscr{Q}$P is an empirical unbiased estimate~$\tilde\varphi$ of the relative unknown phase $\varphi - \Phi$ of a channel
describing an interferometer, with one tunable parameter~$\Phi$ and one unknown parameter~$\varphi$,
typically each being the phase shift of one arm of a two-arm interferometer.
 The imprecision of the estimate~$\tilde\varphi$, namely~$\Delta\tilde{\varphi}_N$,
 is a function of the quantum resource,
 in this case the number~$N$ of particles injected sequentially into the interferometer.
 Without quantum resources,
 the standard quantum limit (SQL) is~\cite{SSW98,BLC92,LBC93,GLM04} 
\begin{equation}
\label{eq:SQL}
\Delta\tilde{\varphi}_N\sim\nicefrac1{\sqrt{N}}.
\end{equation}
The ideal A$\mathscr{Q}$P scheme involves a passive linear lossless two-channel interferometer, such as Michelson, Sagnac or, without loss of generality, Mach--Zehnder~\cite{BW19,YMK86}. 
The interferometer has a pair of input,
and a pair of output,
ports,
which one element of each pair labelled~0 and the other~1,
as depicted in Fig.~\ref{fig:aqp}.
Within the interferometer,
the arms are labelled~0 and~1 as well.
The labelling of output ports conforms to our requirement that a particle injected into the~0 input port emerges from the~0 output port for~$\varphi=0$.
Within the interferometer arm~0 undergoes a known controllable phase shift of~$\Phi$ and arm~1 undergoes an unknown phase shift of~$\varphi$.

Maximal quantum enhancement depends on choosing an appropriate input state~\cite{LSX+20,SP90,BW00},
which we assume to be pure (ideal case),
and is a sequence of photons with the choice of input port a degree of freedom.
Extracting the maximum quantum advantage depends on choosing an appropriate multi-photon input state,
denoted by~$\ket{\Psi}_N$.
This state is a superposition of basis states like~$\ket{01\cdots11}$,
denoting the first photon in port~0,
the second in port~1, the second last in port~1 and the last in port~1, as an example.
Although we do not treat loss here,
we restrict our consideration to  permutationally-symmetric entangled states
(changing the ordering of arrival times does not change the multi-photon state)
as such states are resilient to photon loss~\cite{HS11(2)}.

We now elaborate on the procedure for obtaining~$\tilde\varphi$.
The probability distribution for the unknown phase~$\varphi$ is denoted~$p_N(\varphi;\bm\varpi)$,
with~$\bm\varpi$ labelling a family of distributions,
determined by the input state
(with~$\bm\varpi$ serving as a label for families of input states).
Although the measurement could be general~\cite{SM95}, we restrict measurements to be in the computational basis,
i.e.~single-photon detector at each output port.
The information about single-photon detector clicks is in $b_m \in \{0,1\}$ (0 for detecting at output port~0 and~1 for detecting at output port~1),
with this bit then sent to~C.
Then~C employs policy~$\varrho$,
which we restrict to a binary decision tree~\cite{And97},
to decide from bit~$b_m$ what the next value of~$\Phi_m$ should be. 
After all~$N$ photons have been injected into the interferometer and measurements are completed, 
the final bit string~$\bm{x}_N$ for all measurements maps to~$\tilde\varphi$ according to a formula that is sensitive to the choice of~$\varrho$.

\begin{figure}
\centering
\includegraphics[width=0.35\columnwidth]{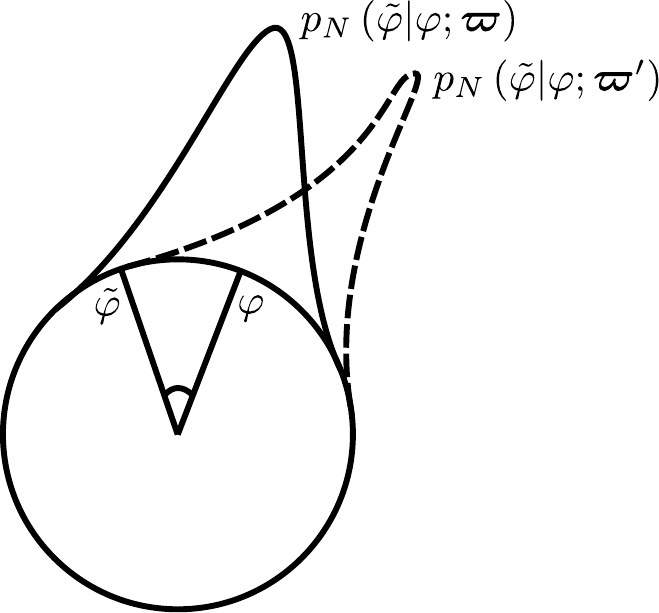}
\includegraphics[width=0.52\columnwidth]{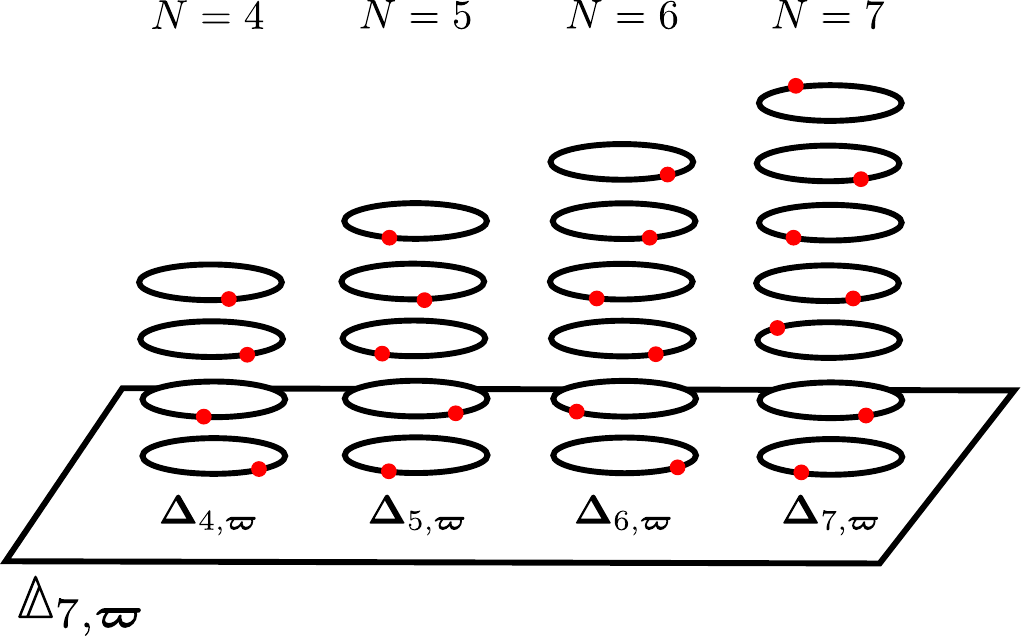}\\
\hspace{-30mm} (a) \hspace{70mm} (b)
\caption{(a) An example of a distribution of the estimate~$\tilde\varphi$, with~$\bm\varpi$ labelling the family of distributions, and~$\varphi$ indicating the value to be estimated. When dealing with distributions over a compact domain, such as a circle, there can be ambiguity in defining covariance. The natural way to address this issue is to embed the circle in a complex plane or represent it by using a set of real numbers. The concept of `sharpness' is introduced to quantify the spread of distribution along the boundary of the circle. The centre of mass, despite the mass being restricted to the boundary, is located inside the circle. The more spread out a distribution is, the further towards the origin the centre of mass goes. For example, for a uniform distribution around the circle, the centre of mass would be at the centre, which would be the most uncertain distribution. The complex number for sharpness represents the coordinate of the centre of mass, with the Fourier transform~\eqref{eq:fouriertransform} yielding the moment-generating function for the distribution. Therefore~\eqref{eq:sharpness} tells us how far away the centre of mass is from the centre without considering the direction.
(b) A pictorial representation of a policy orbit~\eqref{eq:policy_orbit}, with $N_\text{max}=7$, for the A$\mathscr{Q}$P control task. 
Each policy vector~$\bm\Delta_{N,\bm\varpi}$ is represented by the coordinates on the corresponding stack of~$N$ circles, where the red dots mark the values for each component $\Delta_{i,\bm\varpi}$~\eqref{eq:policyN}.
Physically, each red dot in one column represents incremental changes in the estimation of~$\varphi$ with a fixed number of photons~\eqref{eq:update_rule}.}
\label{fig:policy_orbit}
\end{figure}

We now discuss the scaling of $\Delta\tilde{\varphi}_N$~(\ref{eq:SQL})
for A$\mathscr{Q}$P.
~$\Delta\tilde{\varphi}$ is
the standard deviation of~$\tilde\varphi$
if~$\Delta\tilde{\varphi}\in\mathbb{R}$,
but some other natural quantity serving as the variance is needed
given that~$\tilde\varphi$ is a periodic variable; see Fig.~\ref{fig:policy_orbit}(a).
First, we introduce the Fourier transform
\begin{equation}
\label{eq:fouriertransform}
\check{p}_N(\nu;\bm\varpi)
    :=\oint \text{d}\tilde{\varphi}\,
    p_N\left(\tilde\varphi;\bm\varpi\right)
    \text{e}^{\text{i}\nu\tilde\varphi},
\end{equation}
with~$\check{}$ our notation for Fourier transform,
and~$\nu$ is the discrete conjugate variable to phase~$\varphi$.
By setting~$\nu\gets1$,
this Fourier transform~(\ref{eq:fouriertransform}) yields sharpness
\begin{equation}
\label{eq:sharpness}
    S_N(\bm\varpi)
    =\left| \check{p}_N(1;\bm\varpi) \right|^2,
\end{equation}
from which Holevo variance~\cite{WK97, BHB+09},
which is an appropriate variance for a periodic variable,
emerges:
\begin{equation}
\label{eq:HV}
\Delta\tilde{\varphi}_N^2=:V_N:=S_N^{-2}-1.
\end{equation} 
Estimating~$S_N$ is achieved by injecting~$\ket{\Psi}_N$, photon-by-photon, many times for sampling purposes.
The quantum limit to imprecision scaling with respect to~$N$
is the so-called Heisenberg limit (HL)~\cite{BS84}
\begin{equation}
\label{eq:HL}
\Delta\tilde{\varphi}_N\sim\nicefrac1N,
\end{equation}
but this limit is typically not reached in specific operational schemes~\cite{Wis95,WBB+09}.

\subsubsection{Devising feasible control policies for A$\mathscr{Q}$P}
\label{subsubsec:feasible_aqp_policy}

First, we describe relevant elements of $\mathscr{Q}$C.
Then we explain how~$\Phi$ is updated according to~$\varrho$.
Finally, we
introduce the term `policy orbit',
which comprises policies for each $\ket{\Psi}_N$ with $N\in \{4,\dots, N_{\text{max}}\}$.
Finally, we explain the feasibility of policy orbits.

We recast A$\mathscr{Q}$P as $\mathscr{Q}$C, which is expressed in Quotation~\ref{quote:qcontrol1}.
The $\mathscr{Q}$C problem in A$\mathscr{Q}$P falls under the category of measurement-based feedback control, which leads us to describe the components in A$\mathscr{Q}$P as elements of $\mathscr{Q}$C.
~P incorporates the interferometer and its accompanying sensors and actuators.~C is in charge of sending control signal~$u$ to update~$\Phi$.
The update in~$\Phi$ is performed based on a pre-established~$\varrho$ and~P's feedback~$y$, which consists of the photon measurement information bit~$b$ and the current value of~$\Phi$.

\begin{task}[A$\mathscr{Q}$P control task]
\label{task:aqp}
Given an interferometer with an unknown phase shift~$\varphi$ in one arm and a tunable phase shift~$\Phi$ in the other arm,
a quantum state of a sequence of photons~$\ket{\psi}_N$ entering the interferometer, fast measurement of which port the photon leaves and feedback loop to control~$\Phi$, devise a~$\varrho$ for~C that beats the SQL~\eqref{eq:SQL} for estimating~$\tilde{\varphi}$. 
\end{task}

We now discuss how~C adaptively adjusts~$\Phi$ based on~$y$ and~$\varrho$.
Neglecting loss, the $m$th photon comes out from either of the output ports with a probability that depends on $\Phi_m - \Phi_{m-1}$.
We label this outcome by $b_{m}\in \{0,1\}$, where `0' refers to the photon exiting the first port and `1' to the photon exiting the second port.
Given the policy vector
\begin{equation}
\label{eq:policyN}\varrho_{N,\bm\varpi}:=\bm\Delta_{N,\bm\varpi}=(\Delta_{1,\bm\varpi},\Delta_{2,\bm\varpi},\dots,\Delta_{N,\bm\varpi})\in \mathbb{T}^N, 
\end{equation}
for
\begin{equation}
\label{eq:torus}
\mathbb{T}^N:=\varprod^N\mathbb{S},\,
\mathbb{S}:=[0,2\pi)
\end{equation}
being the $N$-torus and the circle coordinate being the angle from~$0$ to~$2\pi$ (radians), respectively.
The value of~$\Phi_m$ is updated sequentially as
\begin{equation}
\label{eq:update_rule}
    \Phi_m = \Phi_{m-1} - (-1)^{b_m}\Delta_m,
\end{equation}
by starting with~$\Phi_m=0$, for every round of measurement~$m$.
Once all photons in the input~$\ket{\Psi}_N$ are exhausted by the $M$th measurement, allowing for the loss of photons such that $1 \le M \le N$, the estimate of $\varphi$ is given by $\Tilde{\varphi} \equiv \Phi_M$. As each measurement outcome~$b_m$ is a binary value, the value of $\tilde{\varphi}$ obtained by this scheme is also discrete.

We now define a policy orbit and its use in assessing scaling.
For a maximum number of photons~$N_\text{max}$, the `policy orbit' is
\begin{equation}
\label{eq:policy_orbit}
\mathbb{\Delta}_{{N_{\text{max}}},\bm\varpi}
:=\left\{\bm\Delta_{N,\bm\varpi}\right\}^{N_\text{max}}_{N=4}
\end{equation}
with the subscript 
\begin{equation}
\label{eq:aqp_pair}
(N_{\text{max}},\bm\varpi)    
\end{equation}
describing both the maximum allowed~$N$ and the input state as a function of~$N$; see Fig.~\ref{fig:policy_orbit}(b).
A feasible policy orbit~$\mathbb{\Delta}^\text{feas}_{N_\text{max},\bm\varpi}$ satisfies~$\Delta\tilde{\varphi}_N$ having approximate power-law scaling with respect to~$N$, 
with the scaling surpassing~$\nicefrac1{\sqrt{N}}$.
Numerically, the feasibility condition can be tested from a log--log plot of $V_N$ vs $N$~\cite{PS19}. 
Specifically, this plot should fit a straight line
\begin{equation}
\label{eq:feasible_porbit}
\log V_N = -2\wp \log N + \text{const},\;
\wp > \nicefrac{1}{2},
\end{equation}
with a goodness-of-fit $\overline{R^2}$ exceeding some acceptable value, e.g.~0.999~\cite{Pal19}.

\subsubsection{Learning for A$\mathscr{Q}$P}
\label{subsubsec:learning_for_aqp}

Now, we critically review existing literature on the topic of ML for A$\mathscr{Q}$P.
We begin by summarizing how meta-heuristic optimization methods have been employed to solve the A$\mathscr{Q}$P task conventionally. We then outline claims to use ML for solving the A$\mathscr{Q}$P task.
Next, we present our criticism of literature casting A$\mathscr{Q}$P as ML based on Mitchell's definition of learning in Quotation~\ref{quote:clearning1} and Sutton's comment on evolutionary methods for solving RL problems in Quotation~\ref{quote:rl_and_evolution}.

A common tool for solving the A$\mathscr{Q}$P control task, namely, Task~\ref{task:aqp}, is to employ meta-heuristic optimization~\cite{ESN+21}.
As examples, particle swarm optimization~\cite{HS10,HS11}, differential evolution~\cite{LCPS13,PS19,PWZ+17}, genetic algorithms~\cite{RDV+20} and Bayesian optimization~\cite{BW00,BWB01} have proven to be useful for Task~\ref{task:aqp}.
Particle swarm optimization was leveraged to estimate $\mathbb{\Delta}^\text{feas}_{N_\text{max},\bm\varpi}$ numerically for Task~\ref{task:aqp}~\cite{HS10,HS11}, and this estimation was employed successfully in an optics experiment~\cite{LPR+18}.
Using differential evolution for A$\mathscr{Q}$P was shown to have a run-time advantage over the above-mentioned technique~\cite{LCPS13}.

In principle, ML can be employed to solve Task~\ref{task:aqp} as well. 
However, claims to use ML for Task~\ref{task:aqp}~\cite{HS10,HS11,PWS16,PWZ+17,RDV+20,COSP21,LCPS13,LPR+18,VPP+20,GGB20,PF20,YTW+20,CVP+23,FSB21,XHFZ19,SLP+19} do not comply with Mitchell's definition of ML stated in Quotation~\ref{quote:clearning1}
or Sutton's criteria for RL in Quotation~\ref{quote:rl_and_evolution}.
Our criticism is restricted to applying ML to Task~\ref{task:aqp}, not to more general instances of using ML for phase estimation~\cite{GFH+17,ZYY+22,NSP21,NPS21,XLL+19,GSG+23,SFB20,KLFM23}.
The reason for our criticism is that each article claiming to employ ML for Task~\ref{task:aqp} actually cast the problem in the context of optimization.

Now we consider specifically RL for Task~\ref{task:aqp}~\cite{HS10,HS11,PWS16,PWZ+17,CVP+23,FSB21}. The criteria for claiming RL is more stringent because both Quotation~\ref{quote:clearning1} and Quotation~\ref{quote:rl_and_evolution} have to be considered.
Specifically, evolutionary algorithms, such as genetic algorithms, genetic programming and simulated annealing, do not qualify as RL because these algorithms do not incorporate learning while interacting with the environment during the lifetime of the RL agent.
Two examples of rejecting membership as RL for Task~\ref{task:aqp} are given by Quotations~\ref{quote:not_rl_aqp_1} and~\ref{quote:not_rl_aqp_2}.
\begin{quo}[Cimini et al.~\cite{CVP+23}]
\label{quote:not_rl_aqp_1}
The method that we chose for the RL algorithm is the cross-entropy method (CEM), which is one of the most generic and easy-to-implement methods. It maximizes the agent’s reward with a derivative-free optimization approach.
\dots\
For this reason, such a method is also called an evolutionary algorithm, since it samples the NN weights from a distribution that is updated at each iteration.
\end{quo}
\begin{quo}[Fiderer, Schuff \& Braun~\cite{FSB21}]
\label{quote:not_rl_aqp_2}
Note that
the cross-entropy method for discrete actions is a reinforcement
learning algorithm [40], while the cross-entropy
method for continuous action spaces, which we use in this
work, is despite its similar name an evolutionary strategy.
\end{quo}

\section{Framework for learning and control}
\label{sec:framework}

Now that we have completed our survey of key background notions and state-of-the-art,
we proceed to set out how we approach building a framework for learning and control, whether the system is appropriately described classically or quantumly.
To begin with, we explain our approach for combining notions of learning and control without assuming underlying classical or quantum constructs.
In other words, we describe how we construct a framework for control whose language does not presuppose either classical or quantum rules.
We then explain how we bring learning into this framework for control.

\subsection{Unifying $\mathscr{C}$C and $\mathscr{Q}$C}

Now, we explain our approach to unifying classical and quantum control by formulating control in a way that is independent of whether the underlying physics is classical or quantum.
First, we introduce a
definition of control
that is agnostic
(literally, `not known')
with respect to a classical vs.\ a quantum framework:
we can incorporate solely classical or solely quantum components (elements of the control system) or a hybrid version with both classical and quantum components.
Second, we explain how we construct a mathematical underpinning for our agnostic definition of control.
Finally, we discuss the elements of our control system, including controller~C, plant~P and communication channels.

In~\S\ref{subsec:control}, we gave two authoritative definitions of a control system, as depicted in Fig.~\ref{fig:control_back}(a).
However, each of these definitions is vexing.
\begin{itemize}
    \item Quotation~\ref{quote:dorf_ccontrol}
defines a control system by how it is made and whether the task is completed but lacks a precise explanation of what must be done well.
\item Quotation~\ref{quote:rosolia_ccontrol}
is unclear regarding what ``sensed quantities'' are and why autonomous behaviour is needed (why cannot a sentient being, like a traffic constable, be incorporated into a control system?).
\end{itemize}
Instead, we define the control system in two ways.
One way is the top-down definition.
\begin{definition}[Top-down definition of control system]
\label{def:top-down}
A control system steers system variables so that pertinent observables reach specific targets.
\end{definition}
\noindent
Here,
the control system is a black-box (the inside is unimportant, hence not seen),
but input, output and performance measure are clear.
The alternative bottom-up definition establishes the control system in terms of its key constituents.

Before providing the bottom-up definition, we define essential components of the control system agnostically. The controller is described by the policy~$\varrho$~(\ref{eq:policy_orbit}),
and the plant is described by a physical process~(\ref{eq:plantdynamics}),
but these two descriptions assume classical input and classical output so are not yet agnostic.
To be fully agnostic,
we consider generalizations of policy and plant dynamics that accept both classical and quantum input and yield both classical and quantum output.
Therefore,
in concordance with Remark~\ref{remark:card}
we redefine error signals~$D$,
actuator actions~$U$ and plant-based sensor outputs~$Y$ as being spaces and do not make suppositions about them being Hilbert spaces or probability spaces.

\begin{definition}[Evaluator]
An evaluator~E is a channel described by $r$-dependent policy~$\varepsilon_r:Y\to D$.
\end{definition}

\begin{definition}[Controller]
\label{def:controller}
A controller~C is a channel described by $r$-dependent policy~$\varrho_r:D\to U$.
\end{definition}
\noindent
This definition captures the essence of Eq.~(\ref{eq:policy})
but does not presuppose that~$D$, $U$
and~$Y$ are classical,
although we treat the reference~$r\in\mathbb{Z}$
and~$\varrho$ as classical.

\begin{remark}[]
Our definition of the controller as an agent does not include sensors or actuators explicitly,
in contrast to typical definitions;
for example,
Quotation~\ref{quote:agent} demands that~C ``can be viewed as [using] sensors and \dots\ actuators'',
but our definition of the full control system
in Definition~\ref{def:top-down}
makes equivalent having sensors and actuators at~P or at~C.
\end{remark}

\begin{remark}[]
We assume that~C is able to read and act on all input data and that all outputs from~C are faithfully sent without loss to~P.
\end{remark}
We separate the agent executing the policy from sensors and actuators,
which we regard as more conveniently defined as part of the plant.
The reason for defining~C this way becomes evident as we describe our whole system-of-system approach to control and learning,
but the short version is that explicit communication lines between~C, P, and eventually the teacher are described better using Definition~\ref{def:controller},
which accepts sensor input and yields outputs sent to actuators.

The controller controls a plant, which we now define.

\begin{definition}[Controlled plant]
\label{def:controlledplant}
A controlled plant~P performs a given task~$\varsigma$,
which maps resources~$R$ 
and controlled actuator input~$U$
to output~$Y$. 
\end{definition}

\begin{remark}[]
This definition for~P does not require~P to be successful at the given task;
the C's job is to steer the plant to the successful execution of the task.
\end{remark}

\begin{figure}[t!]
\centering 
\includegraphics[width=0.4\columnwidth]{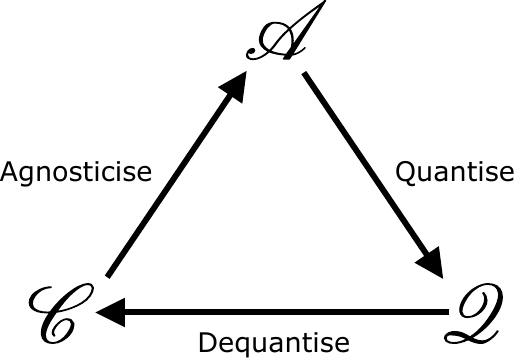}
\caption{A diagram showing the relationships between classical~($\mathscr{C}$), quantum~($\mathscr{Q}$) and agnostic~($\mathscr{A}$) descriptions of a physical system. 
To define any component (e.g.~switch) of our framework, we begin from its~$\mathscr{C}$ description and  ``agnosticise" to construct an equivalent~$\mathscr{A}$ description.
By ``quantising" this~$\mathscr{A}$ description, we then derive the~$\mathscr{Q}$ definition, which can be related back to the original $\mathscr{C}$ description via ``dequantising".}
\label{fig:agnosticise}
\end{figure}

We now discuss the control switch~S.~S is a channel that guides~y to~z if $y=r$ and to~y otherwise; i.e.,
\begin{equation}
\textbf{if}~y=r~\textbf{then}~\text{y}\to\text{z}~\textbf{else}~\text{y}\to\text{y},
\end{equation}
with~$\to$ referring to `guided'. 
In the classical case, the statement~$y=r$ is randomly assigned \textsc{true} or \textsc{false} with the proportion of \textsc{true} vs.~\textsc{false} determined by a threshold value. In the quantum case, $y=r$ is assigned \textsc{true} or \textsc{false} following the quantum fingerprinting technique with a one-sided error~\cite{BCWW01}.
In the following, we first describe a classical~S. Then we agnostisise
(Our term for modifying the description to be equally applicable to quantum and classical cases)
the definition of~S. Finally, we show how the quantum~S yields classical~S in the classical limit; see Fig.~\ref{fig:agnosticise}.
\noindent
\begin{remark}[]
We follow quantum-pseudocode conventions, where 
quantum data and quantum logical operations are distinguished by an underline~\cite{Kni96}.
For example,
\begin{equation}
\underline{\textbf{if}}~\underline{c}~\underline{\textbf{then}}~\text{cnot}(\underline{b},\underline{a}),
\end{equation}
implements a Toffoli gate~\cite{Kni96}, where~$\underline{a}$, $\underline{b}$ and~$\underline{c}$ are quantum registers and $\textsc{cnot}(\underline{b},\underline{a})$ is the controlled-not operation, controlled by the first argument.  
We note that the distinction between the classical and quantum syntactic annotation (meaning the distinction between whether the logic or data are classical or quantum) is primarily semantic and the quantum pseudocode without annotation makes sense operationally as emphasized in the following two quotes.
However, in Definition~\ref{def:quantum_s} of the quantum~S we annotate data and logic by underlining for clarity.
\end{remark}

\begin{quo}[Knill~\cite{Kni96}]
In principle, one can write quantum pseudocode without using annotation.
Note that only registers declared as bit sequences can be used for quantum operations. From an operational point of view it suffices to describe what happens to a register which is currently in superposition when subjected to a classical (non-reversible) operation.
\end{quo}

\begin{quo}[Knill~\cite{Kni96}]
The conventions used here require that a register symbol is always considered either classical or quantum. Semantically, which is in effect depends on the most recent operation applied to it. If it has been declared as quantum, or a proper quantum operation has been applied, then no further classical operations can be used until it is measured. The syntactic annotation helps keep track of the semantics of a register in any given section of code.
\end{quo}

We cast~S in the language of information theory, specifically as a logically reversible gate.
Logical reversibility connects well the classical and quantum descriptions
with reversible logic recasting irreversible Boolean functions as logically reversible permutations whose smallest logical element has three inputs and three outputs~\cite{AGS15}.

\begin{figure}[t!]
\centering
\includegraphics[width=0.85\columnwidth]{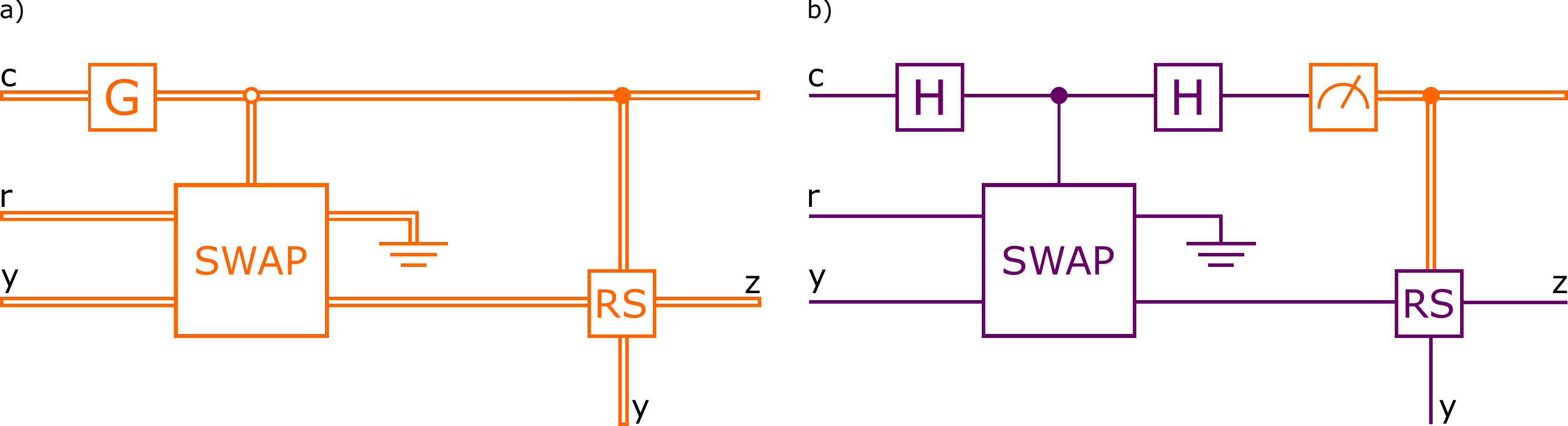}
\caption{A block diagram showing the internal components of a control switch~S, which we have previously represented as a filled circle in Fig.~\ref{fig:control_back}(a). We regard~S as having three inputs, various conditional information-processing steps, and three outputs.
(a) A classical~S has three classical inputs, namely, a control bit~c whose value is $c\in\{0,1\}$, plant-based sensor output dit~y and reference dit~r, and three outputs. The operation~G converts~c into a random Bernoulli variable. If this random variable is~1, r and~y are swapped, followed by the trashing of~r. Finally, the railway switch~(RS) directs~y to the output or the feedback based on the value of the random variable. The three outputs are~c (unchanged from input), y and~z (the control system output).
(b) Quantum~S is similar to classical~S except that we convert data and logic to quantum data and quantum logic, and the controlled-SWAP operation is realized by the quantized Fredkin gate with conjugation with Hadamard gates on the control qubit. Similar to the classical~S, a quantum~S has three inputs~c, r, and~y, which are qubit, qudit and qudit, respectively. The final step of quantum~S is achieved by measuring the control qubit and feeding the one-bit measurement outcome to RS, which directs the qudit~y accordingly.}
\label{fig:control_switch}
\end{figure}

We regard~S as having three inputs,
various conditional information-processing steps,
and three outputs; see Fig.~\ref{fig:control_switch}(a).
The three inputs are the control bit~c
whose value is $c\in\{0,1\}$, 
the plant-based sensor output dit~y and the reference signal dit~r.
The three conditional information-processing steps are
(i)~swap~y and~r
(i.e., \textsc{swap}(y,r)
\textbf{iff} $c=1$
(\textbf{if}~$c$ \textbf{then} \textsc{swap}(y,r)),
(ii)~discard~r (which might have changed value), and
(iii)~\textbf{if}~$c$ \textbf{then}~y~$\to$~z \textbf{else}~y~$\to$~y.
The three outputs are~c (unchanged from input), y and~z (the control system output).

Classically, the first step of~S can be achieved by a Fredkin gate~\cite{FT82} with input~c replaced by the random Bernoulli variable~$G$, which is a random process labelled by threshold value~$q\in(0,1)$ and expressed as 
\begin{equation}
\label{def:g}
\begin{cases}
0 & \text{if unif[0,1]} \geq q,\\
1 & \text{otherwise},
\end{cases}
\end{equation}
with unif referring to a uniform distribution. The threshold value~$q$ is the upper bound for the distance between~$y$ and~$r$ such that those values are deemed to be approximately equal; i.e.,
\textbf{if}~$\operatorname{dis}(y,r)<q$ \textbf{then} \textsc{swap}(y,r)
Conceptually, $q$ should be an overlap
such as the Bhattacharyya coefficient between two probability distributions~\cite{Bha43}.
The Fredkin gate follows and then~r is discarded.
The final step of~S is realized by the railway switch RS,
~\textbf{if}~$G$~\textbf{then}~y $\to$~z \textbf{else}~y $\to$~y.

Before agnosticising~S,
we generalize~c 
to accept either classical or quantum input and to yield classical or quantum output,
if the input is quantum or classical or the input is quantum,
respectively.
Therefore, following Remark~\ref{remark:card},
we regard the set of all control signals $C:=\{c\}$ as forming a Hilbert space
without explicitly considering whether or not the Hilbert space is restricted to a probability space.
\begin{definition}[Agnostic~S]
\label{def:agnostic_s}
A control switch~S is a channel that executes Procedure~\ref{alg:agnostic_switch}
($|$ means `either').
\end{definition}
\begin{algorithm}[H]
\floatname{algorithm}{Procedure}
\begin{algorithmic}[1]
\Require{
\Statex \texttt{bit|qubit} \textsc{\underline{control}} \Comment{control~$c$}
\Statex \texttt{dit|qudit} \textsc{sensor} \Comment{plant-based sensor output~$y$}
\Statex \texttt{dit|qudit} \textsc{external} \Comment{reference~$r$}
}
\Ensure{
\Statex \texttt{bit} \textsc{control}
\Statex \texttt{dit|qudit} \textsc{sensor}
\Statex \texttt{dit|qudit} \textsc{external} \Comment{control system output~$z$}
}
\Procedure{agnosticS}{\textsc{control},\textsc{sensor},\textsc{external}}
\IfThen{\textsc{control}}{swap(\textsc{sensor},\textsc{external})}
\State discard \textsc{external} \Comment{partial trace quantumly}
\State \textsc{control} $\gets$ \textsc{\underline{control}}
\IfThen{\textsc{control}}{\textsc{external} $\gets$ \textsc{sensor}} \Comment{This is control-dependent guiding}
\State \Return{\textsc{control},\textsc{sensor},\textsc{external}}
\EndProcedure
\end{algorithmic}
\caption{Classical$|$quantum control switch}
\label{alg:agnostic_switch}
\end{algorithm}

We now cast~S as a quantum channel by quantizing the agnostic version in Definition~\ref{def:agnostic_s} and Procedure~\ref{alg:agnostic_switch}; see Fig.~\ref{fig:control_switch}(b).
Quantum~S is similar to classical~S discussed above except that we convert data and logic to quantum data and quantum logic, and the controlled-SWAP operation is realized by the quantized Fredkin gate~\cite{PHF+16,NC10} with conjugation with Hadamard gates on the control qubit~\cite{BCWW01}. 
The final step of quantum~S is achieved by measuring the control qubit and feeding the one-bit measurement outcome to RS. 
RS then executes
~\underline{\textbf{if}}~$c$ \underline{\textbf{then}}~\underline{y} $\to$~\underline{z} \underline{\textbf{else}}~\underline{y} $\to$~\underline{y}.

Formally, we quantize~S in Definition~\ref{def:agnostic_s} by modifying Procedure~\ref{alg:agnostic_switch} through the use of quantum syntactic annotation for quantum data and logical operations.
\begin{definition}[Quantum~S]
\label{def:quantum_s}
A quantum~S is a quantum channel that executes Procedure~\ref{alg:quantum_switch}.

\begin{algorithm}[H]
\floatname{algorithm}{Procedure}
\begin{algorithmic}[1]
\Require{
\Statex \texttt{qubit} \textsc{\underline{control}} \Comment{control~$c$}
\Statex \texttt{qudit} \textsc{\underline{sensor}} \Comment{plant-based sensor output~$y$}
\Statex \texttt{qudit} \textsc{\underline{external}} \Comment{reference~$r$}
}
\Ensure{
\Statex \texttt{bit} \textsc{control}
\Statex \texttt{qudit} \textsc{\underline{sensor}}
\Statex \texttt{qudit} \textsc{\underline{external}} \Comment{control system output~$z$}
}
\Procedure{quantumS}{\textsc{\underline{control}},\textsc{\underline{sensor}},\textsc{\underline{external}}}
\qIfThen{\textsc{\underline{control}}}{swap(\textsc{\underline{sensor}},\textsc{\underline{external}})}
\State discard \textsc{\underline{external}} \Comment{partial trace}
\State \textsc{control} $\gets$ \textsc{\underline{control}} \Comment{Measure control qubit}
\cqIfThen{\textsc{control}}{\textsc{\underline{external}} $\gets$ \textsc{\underline{sensor}}}\Comment{This is control-dependent guiding}
\State \Return{\textsc{control},\textsc{\underline{sensor}},\textsc{\underline{external}}}
\EndProcedure
\end{algorithmic}
\caption{Quantum control switch}
\label{alg:quantum_switch}
\end{algorithm}
\end{definition}

\begin{remark}[]
A consequence of quantum~S is that, after each control-loop cycle, y could become entangled with~r, which causes decoherence in~y. 
\end{remark}

\begin{remark}
We now explain how we dequantise quantum~S. In quantum~S, the Hadamard conjugation of~c transforms it from the z-basis to the x-basis and back and then it is measured. The conversion to the x-basis yields a two-by-two pure density matrix. Classically, this conversion is not allowed because there is only one basis in the classical setting. Therefore, to recover classical~S we diagonalize the pure density matrix as it would make it an equal mixture of zeros and ones. This procedure is simulated by~$G$~(\ref{def:g}) in Fig.~\ref{fig:control_switch}(a). 
Whereas there are two Hadamard gates for conjugation in the quantum case, this effect is simulated by~$G$ in the classical~S and the second Hadamard maps to line. The measurement is not necessary either as~c is classical. 
\end{remark}

Next, we define bottom-up control,
as a contrast to top-down control in Definition~\ref{def:top-down}.
Our bottom-up definition is related to Quotation~\ref{quote:dorf_ccontrol},
wherein a control system is described as interconnected components required to build a standard control system.
\begin{definition}[Bottom-up definition of control system]
\label{def:bottom-up}
A control system comprises a controller,
a controlled plant, communication lines between the controller and the plant, input of reference information for establishing targets and output being a task-relevant description of the plant's final state.
\end{definition}
\begin{remark}[]
Two examples of task-relevant descriptions include a simple beep that conveys the plant's task is complete or, alternatively,
sensor readings that convey temperature and consumed energy plus a beep that says the task is completed.
\end{remark}
\begin{remark}[]
Both Defs.~\ref{def:top-down} and~\ref{def:bottom-up} are agnostic with respect to whether the controller, communication channels and plant are each quantum or classical.
\end{remark}

\begin{definition}[Quantum control]
\label{def:qcontrol}
$\mathscr{Q}$C is $\mathscr{C}$C with at least any one of the following being quantum:~E, C, P or~S.
\end{definition}

\begin{remark}[]
Our Definition~\ref{def:qcontrol} reduces to Quotations~\ref{quote:qcontrol1} and~\ref{quote:qcontrol2} subject to the restriction that~P's dynamics are strictly quantum.
\end{remark}

\begin{figure}
\centering
\includegraphics[width=0.7\columnwidth]{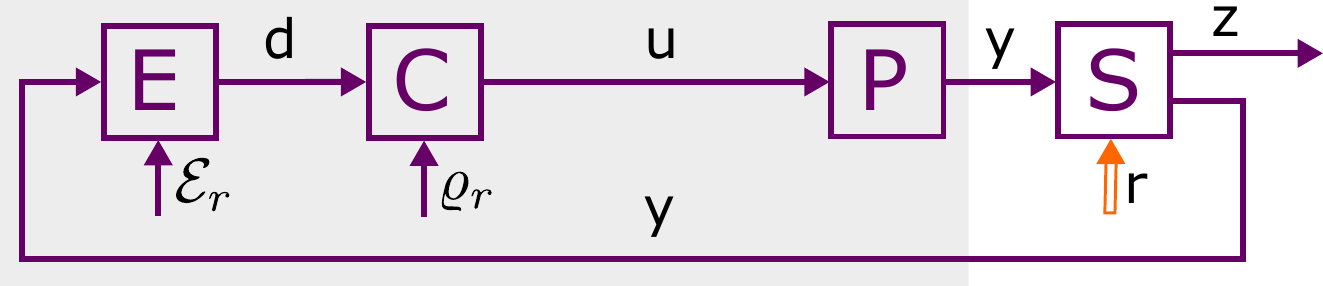}
\caption{A block diagram representing our agnostic closed-loop control system, where the single-line arrows represent the flow of both classical and quantum information, and the double-line arrows represent the flow of only classical information.
We modify the $\mathscr{C}$C loop of Fig.~\ref{fig:control_back}(a) to add an agnostic S and make E, C and P act as either quantum or classical.}
\label{fig:control_agn}
\end{figure}

Now that we have agnostic definitions of the control system in Defs.~\ref{def:top-down}
and~\ref{def:bottom-up},
we describe mathematically the control system and its components
without being explicit regarding whether particular components behave classically or quantumly; see Fig.~\ref{fig:control_agn}.
To this end,
we treat~E, C, P and~S each as a classical or quantum channel,
and these channels are themselves connected to each other by trivial channels,
which also can be classical or quantum.
Together, the controller and the plant characterizes the control system channel.

The input to the control system is zeros 
(i.e., $y\equiv0$,
with~$0$ denoting a string of zeros),
and the output is labelled~z.
The termination condition for the control system is baked into the policy
(i.e., is written into the code for the policy).
We augment the channels between~C, P and~S with an extra bit that carries a termination signal from~C to~S.
Upon receiving the termination signal, S switches its output from~y to~z.
C uses an internal clock,
whose time can be indicated by integers,
to determine whether the termination condition is met. 

\subsection{Unifying $\mathscr{C}$L and $\mathscr{Q}$L}

We now explain our approach to unifying classical and quantum ML. First, we extend the classical definition of ML in Quotation~\ref{quote:clearning1} to the quantum case.
Next, following the definitions of classical and quantum ML we formulate learning in a way that is independent of whether the~$\mathscr{A}$ or~$\mathscr{E}$ are classical or quantum.

We now extend Mitchell's description of ML in Quotation~\ref{quote:clearning1} to the quantum case. We define  $\mathscr{Q}$ML by whether~$\mathscr{A}$ and~$\mathscr{E}$ are quantum.
Building on the effective definition of a learning agent~$\mathscr{A}$ in Quotation~\ref{quote:agent}, a quantum learning agent is defined by whether any components of the agent are quantum.
Quantization of~$\mathscr{E}$ is superposing
inputs, e.g, superposition of features~$\ket{\text{feature}}$ for UL, superposition of labelled features~$\ket{\text{feature}}\ket{\text{label}}$ for SL and superposition of action-observation~\cite{MUP+22} product states~$\ket{\text{action}}\otimes\ket{\text{observation}}$ for RL~\cite{DTB16}.

As an example of quantum~$\mathscr{E}$ in the context of SL with labelled data, consider a data set comprising images of cats and dogs, each with its respective binary label, i.e.~`0' for cat and~`1' for dog.
In the classical setting, the images are encoded in a space that can be bigger or smaller based on feature-engineering techniques~\cite{ZC18}. The feature representing each cat or dog image is a string of zeroes and ones, with one extra bit added to label it.
The set of labelled features then comprises the SL experiences, which are encoded into a classical register to store.
In the quantum setting, quantum~$\mathscr{E}$ is the superposition of labelled features stored in a quantum memory~\cite{DTB16}.
Unlike classical~$\mathscr{E}$, for quantum~$\mathscr{E}$ the memory size can be independent of the number of experiences due to the capability of superposing quantum information.
Similar to the classical setting, though,
the encoding procedure is not necessarily unique.
We can decide to use a larger quantum memory, which would make resolving the quantum experiences easier because of a larger Hilbert-space angle between the feature vectors. 
Quantum~$\mathscr{E}$ can be converted to classical~$\mathscr{E}$ via measurement.

As a classical agent does not benefit from the richness of quantum~$\mathscr{E}$, we only consider a quantum agent in the context of quantum~$\mathscr{E}$.
Now, following the definition of classical ML in Quotation~\ref{quote:clearning1}, we define $\mathscr{Q}$ML as following
\begin{definition}[Quantum ML]
\label{def:qml}
A quantum agent~$\mathscr{A}$ is said to learn from quantum experience~$\mathscr{E}$, with respect to some class of tasks~$\mathscr{T}$ and classical performance measure~$\mathscr{P}$, if its performance at tasks in~$\mathscr{T}$, as measured by~$\mathscr{P}$, improves with quantum experience~$\mathscr{E}$.
\end{definition}

We now introduce a definition of ML that is agnostic
with respect to a classical vs.\ a quantum framework.
Following Defs.~\ref{quote:clearning1} (classical learning)
and~\ref{def:qml} (quantum learning),
plus~Definition~\ref{quote:agent} (agent),
we now define agnostic ML.
\begin{definition}[Agnostic ML]
\label{def:agnostic_qml}
An agent~$\mathscr{A}$ is said to learn from experience~$\mathscr{E}$, with respect to some class of tasks~$\mathscr{T}$ and classical performance measure~$\mathscr{P}$, if its performance at tasks in~$\mathscr{T}$, as measured by~$\mathscr{P}$, improves with experience~$\mathscr{E}$.
\end{definition}

\subsection{Introducing learner and teacher/user}
\label{subsec:learner_teacher}

Having established the unified control and learning schemes, we now introduce the two remaining components of our LfC framework, namely the learner and teacher/user.
To this end, we discuss why we treat the learner as an agnostic agent, whereas we treat the teacher/user as a classical agent.
Additionally, we explain the connection between learner and teacher/user from the perspective of an ML pipeline.

In our framework, we introduce a learner or learning agent~(L) whose purpose is to devise policies for~C to execute. 
We define the combination of~L and~C as a ``learning controller''.
As compared to Fu's definition of a learning controller, our learning controller is not one agent but comprises two separate agents.
This two-agent model simplifies the ideas of quantum and classical learning controllers.
Depending on whether~L and~C are classical or classical four possibilities arise, namely $\mathscr{C}$L--$\mathscr{C}$C, $\mathscr{C}$L--$\mathscr{Q}$C, $\mathscr{Q}$L--$\mathscr{C}$C and $\mathscr{Q}$L--$\mathscr{Q}$C. 
Except for $\mathscr{C}$L--$\mathscr{C}$C, we denote the rest as quantum-learning controllers as they include at least one quantum component.
Nevertheless, our LfC framework needs to be independent of the underlying physics of~C and~L, which leads us to describe~L agnostically.
We represent the learner~L, which executes the learning algorithm, in a purple box to account for the fact that it can be either classical or quantum.
\begin{figure}[t!]
\centering
\includegraphics[width=0.7\columnwidth]{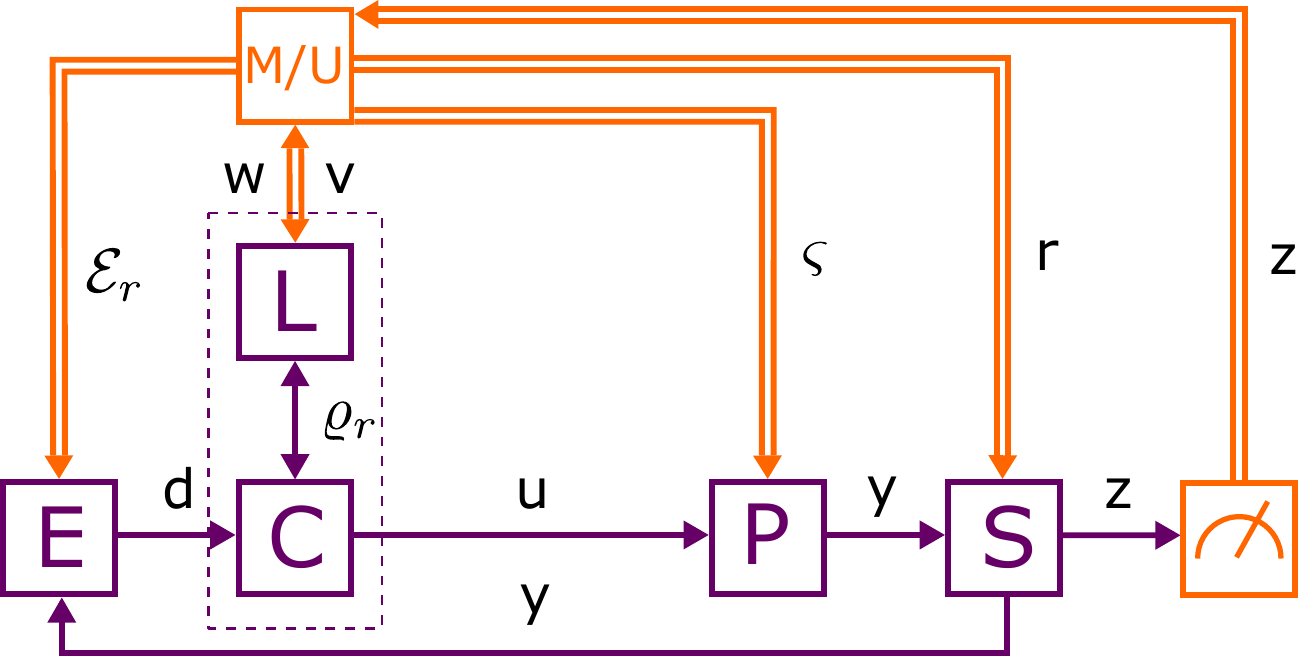}
\caption{A block diagram representing our LfC framework. Double-lined arrows represent the flow of classical information and single-lined arrows represent the flow of both classical and quantum information.
The teacher~M and user~U are classical agents and hence all connections between them and the rest of the blocks are represented by double lines.
We define the combination of~L and~C as a ``learning controller''.
In the training phase, M directs the learning agent~L to devise the control policy~$\varrho_r$ based on~z from~P. M evaluates the performance of C and directs the learning process such that the control system's
overall performance is gradually improved. Additionally, M also sets parameters $\zeta$ of P and provides~r to~S and policy~$\varepsilon_r$ to E. 
In the test and validation phases,~U applies the learned policy~$\varrho_r$ to execute the control task.} 
\label{fig:result2}
\end{figure}

Thus far, our framework requires the policy to appear out of thin air, whereas a complete picture would involve a teacher, whom we denote by~M for the Latin term magister.~M trains~L to develop a policy that becomes useful for the user~U, who replaces~M when the training and validation are complete.  
M is an agent who implements the process of learning for control. Our description of~M is different from Fu's notion that~M is only there to supervise or train.
Our notion of M/U is motivated by considering the physical realization of the ML pipeline~(\ref{eq:ml_pipeline}).

\subsection{Learning for control pipeline}
\label{subsec:mlc_pipeline}

In this subsection, we present our agnostic learning for control scheme in \cref{fig:result2}, where we draw all the  components, namely C, P, S, E, L, M/U, and their interconnections,
and explain how the scheme works.
We begin by describing the data preprocessing, followed by model calibration for our ML for control scheme.
Then, we explain the training and testing of these models.
Finally, we explain the subtleties related to online vs.\ offline learning for control.

We now describe the pre-processing step of the ML pipeline for control.  
In the pre-processing step, the data set
\begin{equation}
\label{eq:mlc_d}
\mathscr{D}=\mathscr{D}_\text{model}\sqcup \mathscr{D}_\text{test}
\end{equation}
is formed by applying pre-processing operations, such as data cleansing, feature extraction and feature selection, on the raw data set~\cite{AOOD20}. 
The raw data set is either generated using a simulation of the control system or created based on data collected after executing the control system without the learning loop for some limited settings.
The pre-processing step usually involves domain expertise and subjectivity~\cite{DCAS22,Huy22}, which we are not analyzing here.

The next step of the ML pipeline is calibrating, which we now describe in detail. In this step, a tuple of feasible hyperparameters of the model is obtained by searching over the hyperparameter space.
During this step, we treat~M and not~U.
The calibrating step involves two sub-steps, namely, training and testing.
In the training step, M obtains a tuple of hyperparameters, a randomly-sampled subset 
\begin{equation}
\label{eq:mlc_dtrain}
\mathscr{D}_\text{train} \subset \mathscr{D}_\text{model}    
\end{equation}
 and the subset
\begin{equation}
\label{eq:mlc_dtest}
\mathscr{D}'_\text{test}=\mathscr{D}_\text{model}-\mathscr{D}_\text{train}
\end{equation}
for training~L. Next, M provides~L with~$\mathscr{D}_\text{train}$ and the hyperparameters and instructs~L to obtain a~$\varrho$ that maximizes the control system's performance on~$\mathscr{D}_\text{train}$.
Then, in the testing step, M provides~L with~$\mathscr{D}'_\text{test}$ and instructs~L to evaluate the performance of obtained~$\varrho$ on $\mathscr{D}'_\text{test}$. 
M and~L collaborate to evaluate this performance for each tuple of hyperparameters by repeating these two sub-steps for different $\mathscr{D}_\text{train}$. After repeating this process of evaluating performance for all possible tuples of hyperparameters, the calibration step returns the tuple corresponding to the maximum~$\varrho$ performance.

In the training step, M provides~L with~$\mathscr{D}_\text{model}$, along with the hyperparameters returned from the calibration step.~M, then instructs~L to obtain~$\varrho$ on this subset. Finally, in the testing step, M instructs~L to provide~C with the~$\varrho$ obtained from the training step.~M proceeds with providing~C with~$\mathscr{D}_\text{test}$, which is unseen in the calibrating and training steps, and instructs~C to commence the control loop.~L remains inactive for the remainder of the time.
The control system executing~$\varrho$ either passes or fails at the test step; if the control system passes, it is then used in the real world, wherein~M is replaced by~U~\cite{Huy22}.

Here, we discuss the subtleties of training and testing a learning controller in both online and offline settings. Online ML methods are employed when~$\mathscr{E}$ become available in sequential order and at each step of the control process. In online learning, there is no distinction between the training and testing stages. Therefore, the roles of~M and~U become identical. In this setting, M/U instructs both~L and~C to interact with~P to generate~$\mathscr{E}$ online.~L then obtains a~$\varrho$ that improves the control system's performance with increasing~$\mathscr{E}$.
On the other hand, offline ML methods obtain a~$\varrho$ for a given data set~$\mathscr{D}$. 
Whereas, in the online learning setting,
$\varrho$ updates as more~$\mathscr{E}$ comes in;
in contrast, offline learning~$\varrho$ is fixed once the training and testing steps are finished. 
In offline learning, C and~P are inactive in the training and testing steps.~M and~L collaborate through the training and testing steps until~L obtains a~$\varrho$ that satisfies the control system's performance requirement. If no~$\varrho$ satisfies the performance requirement, then the control system fails.

\section{Graph representation of state-of-the-art}
\label{sec:graphrep}

In this section, we provide a graphical interpretation of the research done in the field connecting learning and control.
We explain how we identify peer-reviewed literature on any pair of the four topics $\mathscr{Q}$L and $\mathscr{C}$L or $\mathscr{Q}$C and $\mathscr{C}$C.
We employ a square~$\square$ to represent the state of the art with vertices representing each of the four topics and edges representing overlaps between pairs of topics.
Then, we discuss how edges are labelled;
these edge labels represent state-of-the-art.

\subsection{Aggregating candidates}

We begin with explaining how we aggregate the candidates for deciding their memberships as vertices or edges of our knowledge graph. Our collection of candidates is achieved by first searching the literature for relevant peer-reviewed articles. Then we sort these candidates into classes corresponding to different vertex and edge types.

We aggregate relevant literature by employing Google Scholar using $\mathscr{C}$C, $\mathscr{Q}$C, $\mathscr{C}$L and $\mathscr{Q}$L keywords and their pairwise combinations as search prompts. 
The search prompts we obtain either corresponds 
to the four vertices of the~$\square$ or the edges connecting these vertices.
Furthermore, we include references from review articles~\cite{TCC+21,KBC+22,MFS19, LXZ+21,Pre18,MKAL23,ZJQ23,TKP23,GSG+23,DP23,KLFM23,Pez23,MUP+22,GWR22,ML22,BCK+22,CVH+22,DAR+22,HAEM22,DP22,AF21,BC20,CAB+21,BLN+22,AAH+20,PVSS20,Car20,Fu70,WR03,DP10,Rec19,CCC+19,DB18,AT12,KHL+12,HW13,SSP15,ZLW+17,Gor17,CHI+18,BWP+17,MRH18,Lev18,Bar94} to our list of literature.
We then filter out non-relevant articles based on the abstracts and criteria explained in the following subsections~\cref{subsec:graph_vertices} and~\cref{subsec:graph_edges}. 

We then sort the literature into four bins corresponding to the vertices of the~$\square$. One work can be sorted into more than one bin. Next, we assign each article to an edge or hyperedge based on its membership to vertices. If the article is a member of two vertices, we assign it to either a directed or an undirected edge,  which represents using one topic to address another or the unification of topics. If the article is a member of three or more vertices, we assign it to a hyperedge. However, in our search, no candidate was a member of more than two vertices. Therefore, a graph is sufficient to represent our knowledge graph.

\subsection{Vertices}
\label{subsec:graph_vertices}

We now explain the vertices of the~$\square$ and their membership criteria.  Each vertex corresponds to one of the four topics: $\mathscr{C}$C, $\mathscr{Q}$C, $\mathscr{C}$L and $\mathscr{Q}$L. 
We orient the~$\square$ such that the two vertices at the top represent learning,
and the two at the bottom represent control. 
The left side represents the classical regime,
and the right side represents the quantum regime.
We decide membership based on the definitions we have provided for each of these four topics, which in some cases are quotations from the literature and in other cases, they form definitions that we have constructed. 

We now proceed to explain our criteria for deciding whether a given article is accepted as being a member of either or both of the two control vertices.
We use the two authoritative definitions of $\mathscr{C}$C in Quotations~\ref{quote:dorf_ccontrol} and~\ref{quote:rosolia_ccontrol} to decide memberships for the $\mathscr{C}$C vertex. 
A candidate article is accepted as a member of $\mathscr{C}$C if its implicit or explicit definition of $\mathscr{C}$C matches any of the above two definitions.
In the quantum setting, we use our own definition of $\mathscr{Q}$C in
Definition~\ref{def:qcontrol} to decide membership for the $\mathscr{Q}$C vertex.

\begin{remark}
Of course, this procedure for deciding membership is somewhat subjective, so we give an example of one case of rejecting membership to clarify how this procedure works.
An example of rejecting membership is given by Lloyd~\cite{Llo00} for the reason that neither implicit nor explicit separation of~E, S and~M/U are given. The control loop evolves into a coherent superposition of~C and~P without a clear distinction between them.
\end{remark}

We follow Quotation~\ref{quote:clearning1} and our extension to that definition,~Definition~\ref{def:qml}, as the basis for defining learning in both classical and quantum domains. Specifically, we use Quotation~\ref{quote:clearning1} to decide the membership of the $\mathscr{C}$L vertex and Definition~\ref{def:qml} to decide the membership of the $\mathscr{Q}$L vertex. Following our definition of  $\mathscr{Q}$L, we deem an article as being a member of $\mathscr{Q}$L by whether~$\mathscr{A}$ and~$\mathscr{E}$ are quantum.

\subsection{Edges}
\label{subsec:graph_edges}

We now discuss the types of edges used to connect vertices in the~$\square$.
An edge represents articles that incorporate both topics represented by the two vertices connected by the edge. 
A directed edge represents articles that use one topic to address another, and an undirected edge represents literature uniting the two topics.
The absence of an edge represents a lack of literature relating these two topics.
We also describe a hyperedge for the cases where an article covers three or more topics.
Each edge is labelled by the list of its members, i.e., references pertaining to that overlap.

Instances of articles in the literature are decided to be members of a vertex or an edge. 
A directed edge membership is decided based on whether the literature only shows how to use knowledge in the vertex at the tail end of the edge for the topic represented by the vertex at the head of the edge.
We refer to the knowledge represented by a directed edge as the knowledge going one way, i.e., from the `tail' topic to the `head' topic.
An edge is directed only if all the literature is only one-way.

We now proceed to explain our criteria for deciding undirected-edge memberships.
An undirected edge membership is decided based on whether the literature is both a member of two vertices and the two directed edges that connect them.
The undirected edge thus shows that the aggregate knowledge in the literature goes both ways, \emph{not} that all the literature goes both ways. 
Undirected edges signify some advance towards unifying the two topics.
We use our top-down and bottom-up agnostic definitions of control, Defs.~\ref{def:top-down} and~\ref{def:bottom-up}, respectively, to decide membership for articles that signify some advance towards unifying $\mathscr{C}$C and $\mathscr{Q}$C,
and our agnostic definition of ML, Definition~\ref{def:agnostic_qml}, to decide membership for articles that signify some advance towards unifying learning in classical and quantum domains.

Edge memberships are useful if the literature strictly connects only two topics at a time.
If an article were to be a member of three or more vertices, we would use a hypergraph with hyperedges connecting two or more vertices together~\cite{Ouv20}.
As we have not found such an article, a graph is sufficient to represent our knowledge graph.
Our article unites all four topics, so would require a hyperedge if it were to be included in the knowledge representation, but 
we do not include our article in this set.

\subsection{Knowledge graph}
We now represent key literature on unifying classical and quantum control and learning as a knowledge graph~\cite{HBC+21}. We employ a square~$\square$ to represent our knowledge graph with vertices representing each of the $\mathscr{C}$C, $\mathscr{Q}$C, $\mathscr{C}$L and $\mathscr{Q}$L topics and both directed and undirected edges representing connections between topic areas. Our knowledge graph is particularly useful to convey not just state of the art but also knowledge gaps. We begin with describing the~$\square$. Then we explain each of the connections in our knowledge graph and identify the knowledge gaps in the literature. 
\begin{figure}[t!]
\centering\includegraphics[width=0.7\columnwidth]{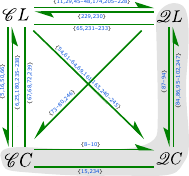}
\caption{Knowledge graph representing key literature concerning the connection between each of the $\mathscr{C}$C, $\mathscr{Q}$C, $\mathscr{C}$L and $\mathscr{Q}$L topics.
Vertices represent each of the four topics, and both directed and undirected edges represent connections between topic areas. The edges are labelled by the list of corresponding member articles.
If an article is a member of two vertices, we assign it to either a directed or an undirected edge, which represents using one topic to address another or the unification of topics.
Edge memberships are useful if the literature strictly connects only two topics at a time. If an article were to be a member of three or more vertices, we would use a hypergraph with hyperedges connecting two or more vertices together.
As we have not found such an article, a graph is sufficient to represent our knowledge graph. An example of a hyperedge is represented by the shaded grey area that connects $\mathscr{C}$C, $\mathscr{Q}$C and $\mathscr{Q}$L.
Our knowledge graph is particularly useful to convey not just state of the art but also knowledge gaps, which are represented by the missing edges.}
\label{fig:knowledgegraph}
\end{figure}

Our knowledge graph,
shown in Fig.~\ref{fig:knowledgegraph},
has four vertices
labelled $\mathscr{C}$L for classical learning,
$\mathscr{Q}$L for quantum learning,
$\mathscr{C}$C for classical control, and
$\mathscr{Q}$C for quantum control.
We do not discuss the status of knowledge for each of these four topics;
rather we are interested only in works establishing connections between these four topics. Our graph differs from usual graphs by having three kinds of edges allowed between vertices simultaneously.

Edges represent connections between pairs of the four topics, with each topic represented by a vertex.
We employ composite edges corresponding to directed and undirected edges.
Each edge in the knowledge graph 
is either unidirectional if one topic builds on the other or undirected if each topic builds on the other.
The essence of the unidirectional $\mathscr{C}$L-$\mathscr{Q}$L edges concerns extensions of $\mathscr{C}$L techniques to $\mathscr{Q}$L and quantum algorithms that enhance classical ML~\cite{DHY+22,OTM+21,SBBE22,BWP+17,WBL12,WCY23,RBN+23,LSS+22,MMG+23,VPB18,WKS15,KA23,LDD20,DTB16,PL23,ASZ+21,Liu2021,DTB17,JFP+23,LSII+20,PDM+14,RML14,NVS+20,SAH+21,SCMB12,HBM+21,LR18,SK19,HCT+19,PBRR18,DBS21} for the rightward direction
and quantum-inspired algorithms for classical ML~\cite{Tan19,ADBL20} in the case of the leftward-pointing edge.
The undirected $\mathscr{C}$L-$\mathscr{Q}$L edge represents literature reporting an advance towards unifying the two topics~\cite{ACL+21,AdW17,HPM+19,GK10}.
The unidirectional $\mathscr{C}$C-$\mathscr{Q}$C edge represents
literature concerning generalization of $\mathscr{C}$C schemes to $\mathscr{Q}$C~\cite{ZLW+17,WM09,Jac14} and the undirected $\mathscr{C}$C-$\mathscr{Q}$C edge represents literature working toward a unifying framework for $\mathscr{C}$C and $\mathscr{Q}$C~\cite{VPS18,PR10}. 
The downward unidirectional edge from $\mathscr{C}$L to~$\mathscr{C}$C represents literature extending~$\mathscr{C}C$ to allow for the controller to learn~\cite{MRH18,HW13,DBN17,Fu70}, and the upward unidirectional edge concerns literature employing  mathematical tools developed for $\mathscr{C}$C to solve problems in $\mathscr{C}$L, such as optimal parameter tuning for training neural networks~\cite{EHL19,LT19} and optimal control techniques for RL~\cite{JL20,BBS95,VPAL09,LXZ+21,SB18}.
The undirected $\mathscr{C}$L-$\mathscr{C}$C edge represents literature reporting an advance towards unifying the two topics~\cite{SBW92,SFPP20,BLN+22,BGH+22}.
The sole $\mathscr{C}$L-$\mathscr{Q}$C edge is unidirectional, and represents articles employing ML for solving problems in~$\mathscr{Q}$C such as quantum-gate design~\cite{PTF20,BCR10,AZ19,NBSN19,SEL+22,BDS+18,MDS+17,NSP21,XLL+19,WEH+16,AN17,Don20,MGCC15,BPB16}.
The sole $\mathscr{Q}$L-$\mathscr{C}$C edge is unidirectional and represents articles employing  $\mathscr{Q}$ML for solving problems in~$\mathscr{Q}$C.
 $\mathscr{Q}$ML for $\mathscr{C}$C has been studied in the context of quantum RL and is limited to solving simple $\mathscr{C}$C problems~\cite{CYQ+20,LS20,LS21,SWIK22,Baq23,SJD22,JGM+21,CLG+18,LCG+17}.
The downward unidirectional edge from $\mathscr{Q}$L to $\mathscr{Q}$C represents literature employing mathematical tools developed for $\mathscr{Q}$L to solve problems in $\mathscr{Q}$C~\cite{HK21,DBC22,SCXC22,LPQ+22,BC21,LJG+23} or $\mathscr{Q}$ML for solving problems in $\mathscr{Q}$C~\cite{WJW+23,SSB23},
and the upward unidirectional edge concerns literature employing  mathematical tools developed for $\mathscr{Q}$C to solve problems in $\mathscr{Q}$L, such as diagnosing barren plateus~\cite{LCS+22} and tuning variational quantum algorithms~\cite{MGB+21,dKTK23,LWC+22,AACA22,IMR+22,MRBA16,YRS+17,CDB+21,MAG+21,WCXL20}.

Now we discuss the missing edges, evident in Fig.~\ref{fig:knowledgegraph}, which represent gaps in the literature.
The first notable gap is represented by the lack of a directed $\mathscr{Q}$C-$\mathscr{C}$C edge. An example of filling this gap would be an article employing mathematical tools developed for $\mathscr{Q}$C to enhance $\mathscr{C}$C.
The gaps represented by the lack of directed $\mathscr{C}$C-$\mathscr{Q}$L and $\mathscr{Q}$C-$\mathscr{C}$L edges are surprising because literature exists that employs mathematical tools developed for $\mathscr{C}$C and $\mathscr{Q}$C to solve problems in $\mathscr{C}$L and $\mathscr{Q}$L respectively. Examples of filling these gaps would be articles employing $\mathscr{C}$C to enhance $\mathscr{Q}$L or presenting quantum-enhanced control leading to superior RL.
The undirected $\mathscr{Q}$C-$\mathscr{Q}$L gap is interesting because literature exists that reports an advance towards unifying $\mathscr{C}$C with deep learning and RL~\cite{SBW92, SFPP20, BLN+22, BGH+22} and the extension to the quantum case follows naturally.

\section{A$\mathscr{Q}$P as a SL problem}
\label{sec:casting}

In this section, we apply our unified LfC framework to describe A$\mathscr{Q}$P as a SL problem. 
First, we motivate and present our idea of elevating the original control task in A$\mathscr{Q}$P to a generalized control problem, which is to obtain~$\mathbb{\Delta}^\text{feas}_{N_\text{max},\bm\varpi}$ for an unknown $p(\varphi;\bm\varpi)$.
Then we discuss the mapping of this control problem to the corresponding learning task, i.e., learning~$\mathbb{\Delta}^\text{feas}_{N_\text{max},\bm\varpi}$.
Finally, we elaborate on our choice for the appropriate learning algorithm.

\subsection{Mapping A$\mathscr{Q}$P to learning}

Here we elaborate on an example of an application of our LfC framework. Specifically, we analyse the A$\mathscr{Q}$P learning problem.
First, we discuss how we elevate the original A$\mathscr{Q}$P control task to a generalized control problem. Then we elaborate on our approach for mapping the generalized A$\mathscr{Q}$P control problem to an ML problem. Finally, we explain our approach for constructing a computationally feasible~$\mathscr{D}$ for our formulation of the A$\mathscr{Q}$P ML problem.

We generalize the A$\mathscr{Q}$P control problem, as described in~\cref{sec:AQEM}, and construct a new control task. For given~$\bm\varpi$,
the re-defined task is to devise~$\mathbb{\Delta}^\text{feas}_{N_\text{max},\bm\varpi}$, defined just below Eq.~\eqref{eq:policy_orbit}, for any unknown $p(\varphi;\bm\varpi)$.
Previously, devising~$\mathbb{\Delta}^\text{feas}_{N_\text{max},\bm\varpi}$ is achieved by obtaining~$\bm\Delta_{N,\bm\varpi}$~\eqref{eq:policyN} for each~$N$ through constrained optimization. The constrained optimization problem, whose fitness landscape is highly non-convex, is solved using heuristic global-optimization algorithms such as particle swarm optimization~\cite{HS10,HS11} and differential evolution~\cite{LCPS13,PS19,PWZ+17} in congruence with the scaling condition~\eqref{eq:feasible_porbit}.
Each of these heuristic global-optimization algorithms are computationally expensive, thus severely limiting achievable~$N_\text{max}$.
To address this drawback of the existing optimization schemes, we propose an ML algorithm to solve the generalized control task in A$\mathscr{Q}$P. 

We now elaborate on our procedure to map the generalized control task of A$\mathscr{Q}$P as an ML problem.
To cast a control problem into a learning problem, we identify the components of a control task as the components of learning, namely, $\mathscr{A}$, $\mathscr{T}$, $\mathscr{P}$ and~$\mathscr{E}$, with the agents being~L, C and~M described in Fig.~\ref{fig:result2}.
For our A$\mathscr{Q}$P learning problem, $\mathscr{T}$ is the control task described in~\cref{subsubsec:adaptiveqem}.~$\mathscr{A}$ comprises~L and~C, where~L devises~$\varrho$ for~C such that~$\mathscr{T}$ is achieved.~M chooses~$\mathscr{P}$ in congruence with~\cref{eq:feasible_porbit}, which is described based on~r~\cite{DB08}. 
Each datum in~$\mathscr{E}$ comprises~$p(\varphi;\bm\varpi)$, $N_\text{max}$   and~$\mathbb{\Delta}^\text{feas}_{N_\text{max},\bm\varpi}$,
for which we seek an efficient representation expressed in the following definition based on our notion of representation in Definition~\ref{def:representation}.
\begin{definition}
\label{def:efficientrepresentation}
A representation~$\tilde{f}_d$ efficiently represents a function~$f$
if the amount of information (e.g., bits) increases no more than polylog$(\nicefrac1{\epsilon})$
for~$\epsilon$, the distance between the function and the representation.
\end{definition}

\begin{remark}
We represent $p(\varphi;\bm\varpi)$
by a truncated cumulant expansion~(\ref{eq:cumulants})
\begin{equation}
    \bm\varpi_\zeta := \left(\kappa_1,\kappa_2,\kappa_3,\ldots,\kappa_\zeta\right) \in \mathbb{R}^\zeta,    
\end{equation}
which is convenient if the representation is efficient.
Each value of~$p(\varphi;\bm\varpi_\zeta)$ is then computed on-the-fly as the value is required,
turning a large space requirement into a slightly longer computation.
\end{remark}

Now we elaborate on how to construct~$\mathscr{D}$~\eqref{eq:Ddisjointunion} for the A$\mathscr{Q}$P learning problem.
~$\mathscr{D}$ of size~$\mathscr N$ is constructed in the pre-processing step of the ML pipeline~(\ref{eq:ml_pipeline}).
In the pre-processing step, M first constructs a set of parameters~$\bm\alpha_\zeta := \left\{\bm\varpi_{1,\zeta},\ldots,\bm\varpi_{\mathscr{N},\zeta} \right\}$
and provides~P with one element of~$\bm\alpha_\zeta$. 
Then~M collaborates with~C to devise~$\mathbb{\Delta}^\text{feas}_{N_\text{max},\bm\varpi_{i,\zeta}}$ by obtaining~$\bm\Delta_{N,\bm\varpi_{i,\zeta}}$ for each~$N$ through optimization. 
After repeating the process of devising~$\mathbb{\Delta}^\text{feas}_{N_\text{max},\bm\varpi_{i,\zeta}}$ for all elements of~$\bm\alpha_\zeta$, $\mathscr{D}$ comprises the data
\begin{equation}
\label{eq:pairs}
\left(\left(N_\text{max}, \bm\varpi_{i,\zeta} \right),\mathbb{\Delta}^\text{feas}_{N_\text{max},\bm\varpi_{i,\zeta}}\right)\,\forall i \in [\mathscr{N}]
\end{equation}
with $\left(N_{\text{max}},\bm\varpi_{i,\zeta}\right)$
described in Eq.~(\ref{eq:aqp_pair}).
Finally, the pre-processing step returns~$\mathscr{D}$.

\subsection{Formalizing the SL problem}

We now formally cast the A$\mathscr{Q}$P control problem as a SL problem.
First, we describe how we classify the A$\mathscr{Q}$P learning problem as one of the SL, UL or RL paradigms of ML, depicted in Fig.~\ref{fig:ml_paradigms}, based on the nature of~$\mathscr{E}$.
Then we introduce the formal learning problem corresponding to the A$\mathscr{Q}$P control problem. 
Finally, we describe the ML workflow for A$\mathscr{Q}$P, which comprises the calibrating, training and testing steps.

We now classify the A$\mathscr{Q}$P learning problem based on the nature of~$\mathscr{E}$. For our A$\mathscr{Q}$P learning problem, $\mathscr{E}$ can be viewed as comprising the pair $\left(N_\text{max},\bm\varpi_\zeta\right)$ in Eq.~(\ref{eq:aqp_pair}) and the corresponding feasible choice~$\mathbb{\Delta}^\text{feas}_{N_\text{max},\bm\varpi_\zeta}$ for the set~$\mathbb{\Delta}_{N_\text{max},\bm\varpi_\zeta}$ in Eq.~(\ref{eq:policy_orbit}).
Therefore, the data structure in~$\mathscr{E}$ naturally fits the SL paradigm of ML, where~$\mathscr{A}$ devises a labelling map~\eqref{eq:sl_map}
\begin{equation}
\label{eq:aqp_sl_map}
f:\mathbb{Z}\times\mathbb{R}^\zeta\to\varprod_{j=4}^{N_\text{max}} \mathbb{T}^j:
    \left(N_\text{max}, \bm\varpi_\zeta\right)\mapsto\mathbb{\Delta}^\text{feas}_{N_\text{max},\bm\varpi_\zeta}.
\end{equation}
The~pair $\left(N_\text{max},\bm\varpi_\zeta\right)$ is typically known as a feature vector in SL, and each~$\mathbb{\Delta}^\text{feas}_{N_\text{max},\bm\varpi_\zeta}$ corresponds to a label.

We now formalize the A$\mathscr{Q}$P SL problem. For a SL problem, we are given~$\mathscr{D}$~\eqref{eq:sl_dataset}, where
each datum is a tuple $(\bm{x}_i,\bm{y}_i)$, with \begin{equation}
\label{eq:xiyi}
\bm{x}_i:=\left(N_\text{max},\bm\varpi_{i,\zeta}\right),\,
\bm{y}_i:=\mathbb{\Delta}^\text{feas}_{N_\text{max},\bm\varpi_{i,\zeta}}.
\end{equation}
The A$\mathscr{Q}$P SL problem then involves devising a labelling  map~$f$~\eqref{eq:aqp_sl_map} such that, for an unseen feature vector $\bm{x}_i$~(\ref{eq:xiyi}), the estimated label is
\begin{equation}
\tilde{\bm y}_i=\tilde{\mathbb{\Delta}}_{N_\text{max},\bm\varpi_{i,\zeta}} \approx \mathbb{\Delta}^\text{feas}_{N_\text{max},\bm\varpi_{i,\zeta}} \in \varprod_{j=4}^{N_\text{max}} \mathbb{T}^j.
\end{equation}
We can use `distance'
\begin{equation}
\label{eq:aqp_loss}
L_{N_\text{max},\bm\varpi_\zeta}:=\text{dist}(\tilde{\mathbb\Delta}_{N_\text{max},\bm\varpi_\zeta}, \mathbb\Delta^\text{feas}_{N_\text{max},\bm\varpi_\zeta}),
\end{equation}
as the loss function to be used during training.
One example of the distance is
\begin{equation}
\label{eq:square_dist}
\text{dist}\left(\tilde{\mathbb\Delta}_{N_\text{max},\bm\varpi_\zeta}, \mathbb\Delta^\text{feas}_{N_\text{max},\bm\varpi_\zeta}\right)
=\sqrt{\sum^{N_\text{max}}_{N=4} \sum^N_{j=1}\left(\tilde{\Delta}_{j,\bm\varpi_\zeta} 
- \Delta^\text{feas}_{j,\bm\varpi_\zeta} \right)^2},
\end{equation}
which is the the square root of the sum of squares of the distances between the label coordinates on each hypertorus~\eqref{eq:policyN},

We now describe the ML workflow for our A$\mathscr{Q}$P SL problem, which comprises the calibrating, training and testing steps. In the calibrating step, M and~L collaborate to obtain a tuple of hyperparameters that correspond to an~$f$~\eqref{eq:aqp_sl_map} which minimizes~$L_{N_\text{max},\bm\varpi_\zeta}$~\eqref{eq:aqp_loss} on a set of randomly-sampled~$\mathscr{D}'_\text{test}$~\eqref{eq:mlc_dtest} according to~\cref{subsec:mlc_pipeline}.  
In the training step, M provides~L with~$\mathscr{D}_\text{model}$~\eqref{eq:mlc_d}, along with the hyperparameters returned from the calibration step.~M, then instructs~L to obtain~$f$ on this subset. Finally, in the testing step, M instructs~L to provide~C with~$\varrho=f$ obtained from the training step.~M proceeds with providing~C with~$\mathscr{D}_\text{test}$~\eqref{eq:mlc_d}, which is unseen in the calibrating and training steps, and instructs~C to commence the control loop.~L remains inactive for the remainder of the time.
The control system executing~$f$ either passes or fails at the test step; if the control system passes, it is then used in the real world, wherein~M is replaced by~U.

\section{Discussion}
\label{sec:discussion}

In this section, we discuss our results.
We begin by discussing our agnostic definitions of control and learning in classical and quantum settings and then interpret our LfC framework.
Next, we analyse the knowledge graph, which we have constructed based on existing literature in the fields of control and learning in both classical and quantum domains. Of particular interest is the identification of gaps, which we found surprising as our approach identifies these gaps and makes it clear that they need further study.
Finally, we explain the relevance of casting A$\mathscr{Q}$P as a SL task.

We have carefully crafted definitions of control and learning that are agnostic in the sense that they hold regardless of whether the underlying physics is classical or quantum.
One key challenge that arises in unifying classical and quantum definitions of control is that the usual notion of a switch in classical control theory no longer applies.
We define a switch which is appropriate for an agnostic approach, Definition~\ref{def:agnostic_s}, in contrast to specifically classical or quantum switches.
Our solution is to cast the classical switch as a logically reversible gate, thus naturally extending the switching operation to the quantum domain.
Interestingly, by following a quantum-pseudocode convention for the quantum switch, we can easily ``switch" between quantum and classical descriptions.
On the other hand, our agnostic definition of learning builds on an established classical ML framework, thus avoiding the discrepancies in the existing definitions of $\mathscr{Q}$ML.

Our main result is our LfC framework which unifies learning and control in both classical and quantum domains.
Our framework includes extending the classical control loop to the quantum case, dealing properly with the switches, controller, what channels are classical, and what are quantum, teacher and the user.
Our framework differs from existing literature in two aspects.
Firstly, in contrast to literature not including a `teacher' and a `user', our framework includes a classical teacher, which trains a learner (classical or quantum) for the control task, and a classical user who replaces the teacher when the training and validation are complete.
Secondly, our work differs from Fu's seminal work~\cite{Fu70} in regards to the separation between controller and learner and the role of teacher.
By treating the learner and controller separately, our framework becomes valid in the quantum domain as we can just make the controller quantum and keep the learner compatible with the classical setting of the real world.
These new features make our framework self-sufficient to be applied to any classical or quantum, or hybrid system.

We present the existing literature on control, learning and their connections, in both classical and quantum domains, in the form of a square graph.
An intriguing fact about this knowledge graph is that it represents only a subset of existing literature in the relevant fields. 
This is because of our strict filtering criteria, which excludes misinterpreted works.
For example, if a paper claims to use ML for a control task but actually uses only optimization under the hood, we exclude that paper.
Thus, in addition to gaps, our knowledge graph also exposes limitations and misinterpretations in the existing literature.

Our unique way of representing literature is particularly interesting because it helps us identify the knowledge gaps, which provides fodder for future researchers.
It is quite surprising to observe that although classical learning is used for both classical and quantum control, the application of quantum learning to quantum control has yet to be fully explored. 
Another interesting observation is that classical and quantum learning has benefitted from developments in classical and quantum control, respectively, but the interrelations between classical control and quantum learning and quantum control and classical learning are not explored yet.
This knowledge graph also conveys that the unification of classical and quantum control (and learning) is not thoroughly explored, which we have addressed in this paper.

Guided by the knowledge of our LfC framework, we address the challenging task of casting A$\mathscr{Q}$P as an
SL problem.
The A$\mathscr{Q}$P control task is computationally expensive. Nevertheless, analysing this task in the light of our framework allows us to employ ML to solve it, i.e.\ to potentially achieve a scaling better than SQL for an unseen unknown-phase probability distribution.
In particular, we first elevate the control in A$\mathscr{Q}$P (Task~\ref{task:aqp}) to a generalized control task amenable to recasting as a SL problem.
By bringing learning into the picture, we could potentially reduce the computational cost for calculating control policies that beat SQL.

\section{Conclusions}
\label{sec:conclusion}

The fields of quantum control and machine learning~(ML) are rapidly progressing, albeit mostly independent of each other.
Although classical control is a widely-popular and established field, quantum control is still in its developing phase.
On the other hand, quantum ML, concerning both quantum-for-learning and learning-for-quantum ideas, is also a very popular research topic.
Despite separate research in these fields, not much attention was paid to unifying the terminologies in quantum control (learning) and classical control (learning); this lack of research might hinder quantum control and learning from fully exploiting established techniques from their classical counterparts.
In this paper, we present unified, i.e.\ agnostic of whether classical or quantum, definitions for control and learning and critically review existing literature in the light of our agnostic definitions.
Moreover, we formulate a learning-for-control framework and explain how supervised learning~(SL) can be used to estimate the unknown phase in a quantum-enhanced interferometric setup.

We review the relevant literature on ML, control systems, unification of classical and quantum mechanics and adaptive quantum-
enhanced interferometric-phase estimation (A$\mathscr{Q}$P).
In particular, we explain key concepts by quoting from authoritative references and, in some cases, formalizing them with mathematical relations.
Moreover, we indicate popular discrepancies in topics, including the definition of quantum ML, evolutionary methods for reinforcement learning~(RL) and the relation between optimization and ML.  
We then discuss the equivalence between classical and quantum descriptions of a physical system based on operational mechanics and geometric correspondence.  
Lastly, we recap the quantum control task in A$\mathscr{Q}$P and the optimization techniques, which were sometimes misrepresented as ML techniques in literature, used to solve this control task.

The main result of our work is a learning-for-control framework, which holds irrespective of whether learning and control are described by classical or quantum mechanics. 
To do this, we first unify classical and quantum control by quantizing the components of a typical closed-loop classical control system, namely the evaluator, controller, plant, switch and their communication channels.
Using a similar approach, we then unify classical and quantum learning.
Finally, we constructed our ``agnostic" framework by elevating an established learning-for-control proposal to account for a classical/quantum description of all components, including the learner, but separating the teacher as a classical agent.

Based on our agnostic learning-for-control framework, we have two more results.
Firstly, we use our unified (quantum and classical) definitions of control and learning to present the existing literature in these fields as a knowledge graph. 
This square graph represents a subset of the existing literature, which has been carefully filtered according to our clarified definitions.
Secondly, we cast the quantum control problem in A$\mathscr{Q}$P as a SL problem and explain the calibration, training and testing steps using our constructed framework.
Although this control problem was framed as an RL problem in the earlier works from our research group, the direct policy search approach does not comply with the conventional RL paradigm as put forward by Sutton.  
Nevertheless, our new way of casting the A$\mathscr{Q}$P problem as a SL problem has potential applications in enhancing quantum clocks~\cite{BS13}
and interferometric position-shift measurements~\cite{Holl79,CTD+80,Cav81}.

Our work leads to many interesting questions and research directions.
Although the field of quantum ML is currently very popular, its applications to classical and quantum control have not been investigated properly.
Another interesting yet unexplored topic is quantum control for classical and quantum learning.
We are particularly excited to study how control strategies on current quantum hardwares can enhance the performance of quantum generative ML to the point of achieving quantum advantage on these noisy devices.

Based on our LfC framework, one can cast the A$\mathscr{Q}$P control task as a SL problem to potentially achieve a scaling better than the standard quantum limit.
Given an unknown phase distribution, obtaining a feasible policy orbit using the existing global-optimization techniques incurs a time-complexity of O($N^6$)~\cite{PS19}, 
which makes the A$\mathscr{Q}$P control task intractable as a new feasible policy needs to be obtained for every unknown phase distribution.
By casting the A$\mathscr{Q}$P task as a SL problem, one can obtain a feasible policy orbit for any unknown-phase distribution in constant time using a pre-trained model.
What remains for future study is to develop efficient SL algorithm to build this predictive model, with the exact computational cost of solving this SL problem dependent on the choice of the algorithm used. 

Beyond the above-mentioned practical applicability, our learning-for-control framework makes the reader ponder more philosophical questions about control, learning and their interconnection. Can M/U be quantum? How does the measurement paradox affect the control and learning feedback loops? Does control theory make sense for a quantum controller? Although we do not have the answers to such questions, one value of our work is that such previously unexplored questions arise very clearly from our framework.

\section*{CRediT authorship contribution statement}
\textbf{Seyed Shakib Vedaie:}
Conceptualization,
Methodology,
Investigation,
Visualization,
Writing -- Original Draft.
\textbf{Archismita Dalal:}
Conceptualization,
Methodology,
Investigation,
Visualization,
Writing -- Original Draft.
\textbf{Eduardo J.~P\'aez:}
Conceptualization,
Methodology,
Investigation,
Writing -- Original Draft.
\textbf{Barry C.~Sanders:}
Conceptualization,
Methodology,
Investigation,
Supervision,
Writing -- Review \& Editing.

\section*{Declaration of competing interest}
The authors declare that they have no known competing financial interests or personal relationships that could have appeared to influence the work reported in this paper.

\section*{Data availability}
No data was used for the research described in the article.

\section*{Acknowledgements}
SSV would like to thank MITACS. AD would like to thank MITACS and the Canadian Queen Elizabeth II Diamond Jubilee Scholarships program. BCS appreciates financial support from the Natural Sciences and Engineering Research Council of Canada. AD, EJP and BCS acknowledge the support of the Major Innovation Fund, Government of Alberta, Canada. We thank Carlo Maria Scandolo and Howard M. Wiseman for useful discussions on quantum feedback.
We acknowledge the traditional territories of the people of the Treaty 7 region in Southern Alberta. 

SSV and AD contributed equally to this work.

\end{document}